\documentclass[superscriptaddress,reprint,nofootinbib,preprintnumbers,amsmath,amssymb,aps]{revtex4-2}
\usepackage{amsmath, mathrsfs, latexsym, amssymb, color, graphicx, comment, hyperref, scrhack}
\usepackage{lmodern}
\usepackage{physics}

\newcommand{\dfr}[2]{\frac {\displaystyle #1}{\displaystyle #2}}

\usepackage{graphicx}
\usepackage{color}
\usepackage[usenames,dvipsnames]{xcolor}

\begin{document}
\title{ Critical Dynamics of the Spin-Boson Model
}
\author{M.\,G.\,Vasin}
\affiliation{N.\,L.\,Dukhov Research Institute of Automatics (VNIIA), 127030 Moscow, Russia}
\affiliation{Vereshchagin Institute of High Pressure Physics, Russian Academy of Sciences, 108840 Moscow, Russia}
\author{S.\,V.\,Remizov}
\affiliation{N.\,L.\,Dukhov Research Institute of Automatics (VNIIA), 127030 Moscow, Russia}
\affiliation{Kotelnikov Institute of Radioengineering and Electronics, Russian Academy of Sciences, 125009 Moscow, Russia}
\affiliation{National Research University Higher School of Economics, 109028 Moscow, Russia}
\author{A.\,A.\,Elistratov}
\affiliation{N.\,L.\,Dukhov Research Institute of Automatics (VNIIA), 127030 Moscow, Russia}

\begin{abstract}
In this work, we study the low-energy properties of the spin-boson model (SBM), which describes the dynamics of a spin-1/2 coupled to a bosonic environment characterized by a power-law spectral density $f(\omega)\propto \omega^s$. The theoretical description is based on the Schwinger--Keldysh technique  combined with a Majorana spinor representation of the spin. This approach enables a renormalization group analysis of the model's critical dynamics without relying on quantum-classical mapping. We show that the transition from a delocalized to a localized state arises due to a Wilson--Fisher fixed point in both the ohmic ($s=1$) and sub-ohmic ($s<1$) cases. Our analysis recovers key results of Leggett's theory and identifies $s=1/2$ as the upper critical dimension, marking the boundary where critical exponents become mean-field. The findings are in good agreement with the predictions of quantum-classical mapping and state-of-the-art numerical data.
\end{abstract}


\flushbottom
\maketitle
\thispagestyle{empty}

\section{Introduction}

The spin-boson model is a ubiquitous model for studying decoherence in quantum two-level systems \cite{RevModPhys.75.715, Nielsen2000}. It has recently attracted significant research attention due to rapid progress in the development of nanoscale devices \cite{RevModPhys.73.357,doi:10.1126/science.1231930,Clarke2008}, which enable the study of quantum system dynamics interacting with an environment.

The spin-boson model describes a two-level system (e.g., a spin-1/2, a qubit, or an atom in a double well) interacting with a surrounding bosonic "bath" (environment). The impact of the bath onto the spin-dynamics is determined by the coupling function $J(\omega)=\alpha\,\omega^s$, a key parameter of which is the exponent $s$. This exponent depends on the properties of the bath and determines the character of the phase transition. The system's behavior critically depends on the low-frequency properties of the bath \cite{RevModPhys.59.1,doi:10.1080/14786430500070396,Weiss2012}, making this model a convenient platform for studying quantum critical phenomena, dissipation, and decoherence.

It is known that at absolute zero temperature, $T=0$, a quantum phase transition (QPT) occurs in the system described by the spin-boson model for $s \leq 1$. This transition is driven by quantum fluctuations (as opposed to thermal fluctuations in classical phase transitions), and the control parameter is the coupling strength, $\alpha$, which characterizes the interaction strength between the spin and the bath. In the weak-coupling regime, the quantum fluctuations induced by the bath weakly influence the spin. Then it can tunnel between the two states (i.e., exist in a superposition of these states) while being in the delocalized phase. Conversely, in the strong-coupling regime, the interaction with the bath suppresses tunneling, localizing the spin in one of its two basis states (e.g., "up" or "down"). In this case, the system is in the localized phase. The transition between these two phases is a quantum phase transition.

As mentioned above, the character of the phase transition depends on the exponent $s$. Three distinct regimes are identified:
\begin{itemize}
\item In the most interesting ohmic regime ($s = 1$), QPT manifests as the transition between the delocalized ($\alpha < 1$) and localized ($\alpha > 1$) phases. It is important that in the delocalized phase at $\alpha=1/2$, a crossover takes place from a damped oscillatory regime ($\alpha < 1/2$) to an incoherent relaxation regime ($\alpha > 1/2$) \cite{RevModPhys.59.1};
\item At the sub-ohmic regime ($s < 1$),  as in the ohmic one, the delocalization--localization transition occurs at a critical coupling strength $\alpha_c$ whose value decreases with $s$ decreasing. This continues until $s$ reaches the value $s=1/2$, below which the critical exponents become mean-field ones \cite{PhysRevLett.91.170601,PhysRevLett.94.070604,doi:10.1080/14786430500070396,
PhysRevLett.102.030601,
PhysRevB.81.121105, PhysRevLett.108.160401,  PhysRevB.85.144425,PhysRevB.100.115106,shen2023numerical}; 
\item At the super-ohmic regime ($s > 1$) the spin is delocalized. 
\end{itemize}

Modern analytical approaches to describing the quantum phase transition in the spin--boson model are based on the quantum-classical mapping method, in which the spin evolution is considered in a kink gas representation, where the kinks represent spin flips along the imaginary time axis. The problem is reduced to the classical one-dimensional Ising chains with isotropic long-range interactions between spins, for which a perturbative RG analysis has been performed by Kosterlitz \cite{kosterlitz1976phase}.
Thus, the transition between localized and delocalized phases in the spin--boson model is described as a Berezinskii-Kosterlitz-Thouless transition \cite{doi:10.1142/6738,doi:10.1080/14786430500070396,PhysRevLett.94.070604}.

{
A wide range of computational tools is currently available for the quantitative investigation of the ground-state properties of the spin‑boson model (SBM). Among them, the most famous one is the Numerical Renormalization Group (NRG) developed by Wilson back in the 1970s \cite{PhysRevB.4.3174}. Interestingly, early implementations of this scheme yielded results that were inconsistent with mean‑field predictions, thereby casting doubt on the universality of the quantum‑classical mapping \cite{PhysRevLett.91.170601,PhysRevLett.94.070604}. However, alternative approaches, such as Quantum Monte Carlo (QMC) \cite{PhysRevLett.102.030601}, Density-Matrix Renormalization Group (DMRG) \cite{PhysRevB.77.174305}, Exact Diagonalization (ED) \cite{PhysRevLett.102.150601}, and Variational Matrix Product States (VMPS) \cite{PhysRevLett.108.160401} demonstrated the opposite. It was later realized that the anomalies in the NRG outcomes stemmed from technical limitations associated with the truncation of the bosonic spectrum \cite{PhysRevB.85.144425}. Then, the quantum-classical mapping results were confirmed using the multiple polaron ansatz in the numerical variational method (NVM) \cite{shen2023numerical}.
}

In this work, we propose an alternative approach to analytically describing the properties of the spin-boson model, based on the methods of critical dynamics.  { We think that since the system is at zero temperature, the low-frequency limit of the dynamic susceptibility directly determines the static response, and the closing of the dynamic gap is equivalent to the condition for the ground-state phase transition.} To implement this approach, we employ a fermionization of the spin using Majorana fermions, building upon the progress made in \cite{PhysRevLett.69.2142,PhysRevB.55.142,PhysRevLett.91.207203,PhysRevLett.91.207204,PhysRevB.93.174420,SCHAD2015401}. These works demonstrate that the correlations between the transverse Majorana fermions can be described by an effective Gaussian action. This allows us to apply standard phase transition theory methods to investigate the model in question. Using a renormalization group analysis, we examine the critical dynamics of the model and derive analytical expressions for the dependence of the critical spin-boson coupling constant and the critical exponent $1/\delta$ on the spectral exponent $s$. The obtained results lead to the conclusion that, in both the sub-ohmic and ohmic regimes, the system undergoes a continuous delocalization--localization phase transition, and the proposed approach provides a comprehensive description of the critical behavior near this transition.

\section{Model}

The spin-boson model involves a single quantum $1/2$-spin $\mathbf{\sigma}$ interacting with the bosonic bath $X$. We assume that only the $x$-th component of the spin interacts. Thus, the spin-boson model Hamiltonian has the general functional form:
\begin{equation}
\mathcal{H}=-\Delta \sigma_z-(\lambda X +h)\sigma_x+\mathcal{H}_X,
\label{H1}
\end{equation}
where $\Delta$ is the tunneling constant (the magnetic field applied to the spin), $h$ is the small ($h\ll\Delta$) field applied along the $x$-axis, which determines the energy bias, $\lambda$ is the bath-spin coupling constant, $X$ is the bosonic field, and $\mathcal{H}_X$ is the Hamiltonian of the bosonic bath. We assume that the bosonic bath is characterized by the spectral function $f(\omega)\propto \omega^s$, where $s$ is the spectral exponent, and $\langle XX\rangle_{\omega}=\mathrm{i}C\,\omega^s$ (see Appendix A). {
The constant $C$ is related to the standard dissipation strength $\alpha$ used in the spin-boson literature \cite{RevModPhys.59.1} by $C = 2\alpha / (\pi \lambda^2\omega_D^{1-s})$, where $\omega_D$ is the Debye cutoff frequency.
}

First, we map the spin-1/2 operator onto fermionic degrees of freedom, which can be done, in particular, with the so-called Majorana fermions \cite{PhysRevLett.69.2142,PhysRevB.55.142,PhysRevLett.91.207203,PhysRevLett.91.207204,PhysRevB.93.174420} (Majorana spinors). This mapping obeys the following correspondence principle:
\begin{equation}
\sigma_{i}=-\frac{\mathrm{i}}{2}\varepsilon_{ijk}\psi_{j}\psi_{k}=\Theta\psi_{i},\qquad i,\,j,\,k=(x,\,y,\,z)
\end{equation}
where $\Theta=-2\mathrm{i}\psi_{x}\psi_{y}\psi_{z}$ ($\Theta^2=1/2$) { is the operator commuting with the Hamiltonian ($[\Theta \mathcal{H}]=0$)}, and $\vec\psi$ is the Majorana spinor field, which obeys the Clifford algebra and has the following properties:
\begin{gather*}
\vec\psi^+=\vec\psi,\quad
\{\psi_{i}\psi_{j}\}=\psi_{i}\psi_{j}+\psi_{j}\psi_{i}=\delta_{ij},\quad
\psi_i\psi_i=1/2.
\end{gather*}
Thus, the Hamiltonian (\ref{H1}) can be written in the Majorana representation as follows
\begin{equation}
\mathcal{H}=\mathrm{i}\Delta\psi_x\psi_y+\mathrm{i}(\lambda X+h)\psi_y\psi_z+\mathcal{H}_{X}. 
\label{Ham}
\end{equation}

{
However, a straightforward perturbative analysis (see Appendix A) reveals that a theory based on the Hamiltonian (\ref{Ham}) is non-renormalizable. This contradicts the observed physical picture, according to which transverse noise acting on the spin merely renormalizes the effective value of $\Delta$. The reason is that in the considered three-Majorana-fermion representation, only two of them are independent. The product of these three fermions, $\Theta=-2\mathrm{i}\psi_{x}\psi_{y}\psi_{z}$, commutes with the Hamiltonian, implying that the eigenvalue of this product is $\Theta = \pm i/\sqrt{2}$, i.e. the fermions are interconnected. One can show this condition ensures that the commutation relations for the spins are satisfied (see Appendix A). In case of the bilinear representation (\ref{Ham}), these commutation relations not are taken into account at the expanding into a perturbation theory series, that does not allow for the construction of a diagram renormalizing $\Delta$.  To overcome this and correctly account for spin commutation relations in the diagrammatic expansion, we rewrite the Hamiltonian in an equivalent but renormalizable form:
}
\begin{multline}
\mathcal{H}=-\Delta\Theta\psi_z+\mathrm{i}(\lambda X+h)\psi_y\psi_z+\mathcal{H}_{X}\\
=\mathrm{i}2\Delta\psi_x\psi_y\psi_z\psi_z+\mathrm{i}(\lambda X+h)\psi_y\psi_z+\mathcal{H}_{X}. 
\label{Ham2}
\end{multline}
Relying on the properties of Majorana spinors, it is straightforward to see that such a representation of the Hamiltonian is physically equivalent to the previous one. However, it now becomes renormalizable, since the form of the vertex $\Delta$ allows us to account for the commutation relations of the spin components (see Appendix A).

\section{Transition between localized and delocalized states}

We investigate the non-equilibrium dynamics of the presented model in terms of the Schwinger--Keldysh technique \cite{Kamenev_2011}. In the one-loop approximation (see Fig.\,\ref{PF2}), the model parameters are renormalized as follows (see Appendix A):  
\begin{align}
{\Delta}' &\approx \Delta-\frac{\Delta\lambda^2}{2}\int_{\omega'}^{\omega_D}\frac{\mathrm{d}\omega}{2\pi}\frac{C\omega^{s}(\Delta^2-\omega^2)}{(\Delta^2+h^2-\omega^2)^2}, \label{R16} \\
{h}' &\approx h-\frac{h\lambda^2}{2}\int_{\omega'}^{\omega_D}\frac{\mathrm{d}\omega}{2\pi}\frac{C\omega^s}{\Delta^2+h^2-\omega^2}, \label{R1} \\
{\lambda}' &\approx \lambda -\frac{\lambda^3}{2}\int_{\omega'}^{\omega_D}\frac{\mathrm{d}\omega}{2\pi}\frac{C\omega^{s}(\Delta^2-\omega^2)}{(\Delta^2+h^2-\omega^2)^2}, \label{Lambd:S}
\end{align} 
where $\omega_D$ is the upper cutoff limit (Debye frequency) and $\omega'$ is the lower limit of integration.    
\begin{figure}[ht]
   \centering
   \includegraphics[scale=0.37]{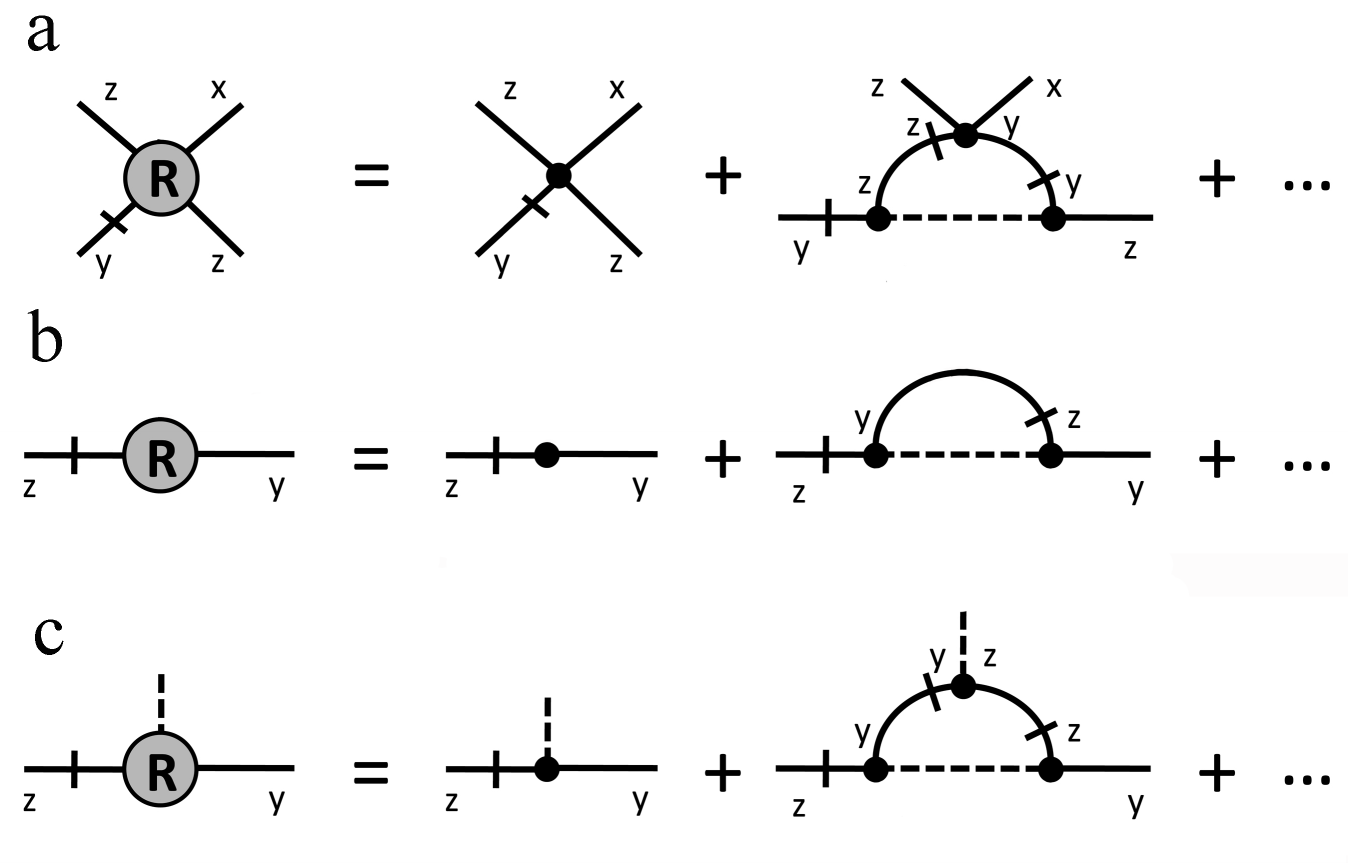}
   \caption{Diagrams contributing to the one-loop approximation of the renormalization of the $\Delta$ (a), $h$ (b), and $\lambda$ (c) vertices. Solid lines correspond to spinor correlators, dashed lines to the boson field correlator (see Appendix A).}
   \label{PF2}
\end{figure}

The critical behavior of the system can be considered within the technique of critical dynamics \cite{Vasil'ev_2004} based on the hypothesis of dynamical scaling. As well as in Leggett's theory, in the approach we are considering, the Ohmic regime proves to be the most convenient for studying the critical transition regime between localized and delocalized states.

\subsection{Ohmic regime}

In the Ohmic case, $f(\omega)\propto \omega$, the theory is logarithmic because the renormalization procedure is independent of the scale (the cutoff parameter)(see Appendix B). Since in logarithmic theory, the low-energy infrared divergences in the critical region coincide with the high-energy ultraviolet ones, this allows us to use the renormalization group (RG) method.  

For the sake of simplicity, we assume that $h$ is very small and can be neglected. Furthermore, lets write the renormalization expressions using the causal propagators, which is allowable for Wilson's RG procedure since it considers the ultraviolet renormalization. Thus, in the one-loop approximation, we write the following renormalization group equations (see Appendix B):
\begin{equation}\label{R4}
\left\{
\begin{array}{l}
\displaystyle\frac{\mathrm{d}\Delta}{\mathrm{d}\ln\Lambda} \approx \Delta\left(1-\frac{\lambda^2C}{4\pi}\right),\\
\displaystyle\frac{\mathrm{d}\lambda}{\mathrm{d}\ln\Lambda} \approx \lambda\left(\frac{1}2-\dfr{\lambda^2 C}{4\pi}\right).
\end{array}
\right.
\end{equation}
from which one can see that the theory contains two fixed points: the Gaussian fixed point, $\lambda =\Delta=0$, and the Wilson--Fisher (WF) fixed point, $\Delta=\Delta^*=0$, 
\begin{gather}
\lambda=\lambda^*=\sqrt{{2\pi}/{C}}.
\label{SS}
\end{gather}
The second one corresponds to a second-order phase transition.

However, in contrast to the fluctuation theory of continuous phase transitions, where the infrared limit of integration over frequency $\omega'\to 0$, our theory posits that this limit is a nonzero cutoff energy. It is due to the fact that the spin cannot absorb or emit energy less than $\Delta$, i.e. $\omega'\to\Delta$. What gives the situation its piquancy is the fact that the cutoff frequency is proportional to the quantity that is being renormalized itself. Therefore, the renormalization equations are analyzed in the adiabatic approximation, which is used in Leggett's seminal work \cite{RevModPhys.59.1}. This approximation assumes that the region of integration over frequency is bounded below by the condition  $\omega>\omega'>p\Delta$ ($p\gtrsim 1$), where $p\Delta$ is the lower cutoff limit, which is small compared to the Debye frequency, $\Delta\ll \omega_D$. As we noted above, each rescaling iteration leads to the renormalization of the lower cutoff of integration, $p \Delta \to p \Delta'$.

\begin{figure}[ht]
   \centering  \includegraphics[scale=0.6]{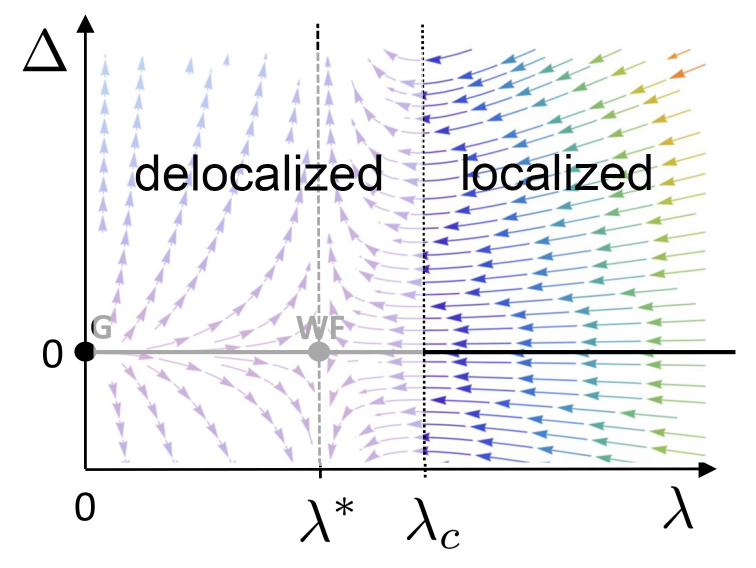}
   \caption{
  Schematic representation of renormalization group flows in the ohmic case. The critical point $\lambda^*$ corresponds to the $\alpha=1/2$ in the Leggett's theory \cite{RevModPhys.59.1}. At $\lambda>\lambda_c\equiv\sqrt{2}\lambda^*$ ($\alpha>1$) the spin is in localized state.}
   \label{STR}
\end{figure}

As a result, there are two renormalization regimes, the boundary between which is determined by the condition
$\mathrm{d}\ln\Delta/{\mathrm{d}\ln\Lambda}=0 $: 
From the first equation in (\ref{R4}), one can see that in the region $\lambda>\lambda_c\equiv\sqrt{2}\lambda^*$, every iteration of the renormalization procedure shifts the lower cutoff frequency down $\Delta\to\Delta'<\Delta $. In this region, the renormalization procedure does not differ from that in the canonical theory of continuous phase transitions, and the iterations do not stop until the renormalized tunneling constant $\Delta$ reaches zero.  However, at $\lambda<\lambda_c$, the iteration process stops when $\Delta'$ reaches some finite value $\Delta_{min}$ \cite{RevModPhys.59.1,PhysRevB.9.215,hewson1980local}. This value, in the ohmic case ($s =1$), can be estimated as 
\begin{gather}\label{REN}
  \Delta_{min}= \omega_D\left(\frac{\Delta}{\omega_D}\right)^{\frac{\lambda_c^2}{\lambda_c^2-\lambda^2}} 
\end{gather} 
(see Appendix C). 
Therefore, if the initial parameters of the system are in the region $\lambda>\lambda_c$, then the renormalized tunneling frequency vanishes, which corresponds to localization.
In contrast, If the initial coupling constant $\lambda<\lambda_c$, then the renormalization of the tunneling frequency leads to its decrease to some finite but non-zero value $\Delta_{min}=\omega_D(\Delta/\omega_D)^a$ (Lamb shift), where $a$ is a positive constant (see Fig.\,\ref{STR}). As a result, the spin is in the delocalized state. 

Thus, the extremum points of the RG flow lines, $\mathrm{d}\ln\Delta/{\mathrm{d}\ln\Lambda}=0$, correspond to the transition from the localized state to the delocalized state. As can be seen from Eq. (\ref{R4}), in the Ohmic case, these extrema lie on the line $\lambda =\lambda_c$. This localization transition is analogous to Anderson's orthogonality catastrophe~\cite{PhysRevLett.18.1049}, where the effective tunneling amplitude is renormalized by the overlap integral, which vanishes for $\lambda>\lambda_c$~\cite{RevModPhys.59.1}. 

Note that the theoretical (UV) WF fixed point is located in the region of $\lambda$ values ($\lambda<\lambda_c$) where the IR limit is bounded by a finite value of $\Delta_{min}$. Therefore, the RG group action in this region is violated. We assume that our system is located in a fluctuation region close to the theoretical fixed WF point, where we can use RG analysis. However, this point itself is physically unattainable due to the violation of the RG group action in the region $\lambda<\lambda_c$.

\subsection{Super-Ohmic regime}

In non-Ohmic regimes, the theory is not logarithmic. As a result, the critical coupling constant depends both on $\Delta$ and $s$. Lets consider the super-Ohmic case ($s>1$). As we have found above, the value of the coupling constant $\lambda_c$ corresponding to the transition between the localized and delocalized states is determined by the condition 
\begin{gather}\label{Condition}
1-\Delta^{s-1}\frac{\lambda^2C}{4\pi}=0,  
\end{gather}
which follows from (\ref{R16}). Since $s>1$, then $\Delta^{s-1}\to 0$ at $\Delta, h\to 0$.
This means the renormalization corrections become negligible, and the parameters $\Delta$, $h$, and $\lambda$ are not renormalized. As a result, $\mathrm{d}\Delta/\mathrm{d}\ln\Lambda>1$ at any $\lambda$; i.e., the system is always delocalized.  
For the same reason, there is no WF fixed point at $\lambda >0$, and there remains only one unstable Gaussian fixed point that corresponds to the delocalized state. Thus, in the Super-Ohmic regime, there is no transition from the delocalized phase to the localized phase; the spin is in a delocalized state at any coupling constant.

\subsection{Sub-Ohmic regime}

\begin{figure}
\centering
\includegraphics[scale=0.7]{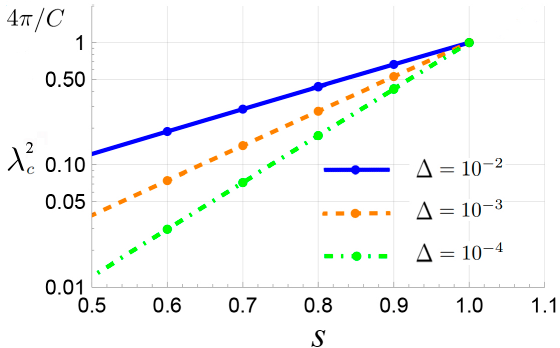}
   \caption{\color{black} The dependence of  $\lambda_c^2$ value on the $s$ exponent.  }
   \label{Lambda}
\end{figure}

\begin{figure}[ht]
   \centering
\includegraphics[scale=0.7]{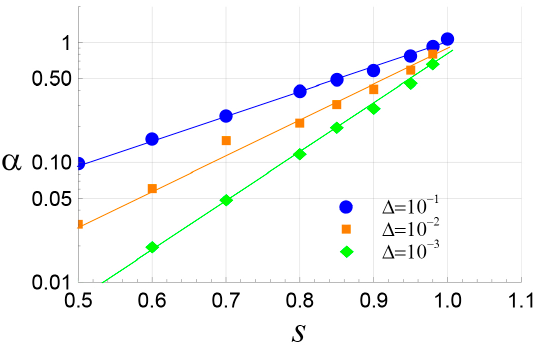}
   \caption{\color{black} The dependence of the critical value of  $\alpha $ (in the our case $\lambda^2\propto \alpha$) on the $s$ exponent in semi-logarithmic coordinates calculated from the numerical results  determined in \cite{PhysRevLett.91.170601} from the numerical renormalization group method.}
   \label{Lambda2}
\end{figure}

In the sub-Ohmic case, the renormalization corrections become relevant, and the above condition for the transition between the localized and delocalized states (\ref{Condition}) indicates that the dependence of the critical coupling parameter on the exponent $s$ follows a power law 
\begin{gather}\label{LL}
\lambda_c^2(s)\equiv{{4\pi \Delta^{1-s}}/{C}},
\end{gather}
which is consistent with known numerical findings \cite{PhysRevLett.91.170601} (see  Fig.,\ref{Lambda} and Fig.,\ref{Lambda2}).  One can see that the transition point between localized and delocalized states $\lambda_c$ exponentially decreases as $s$ decreases, which shows the narrowing of the delocalization region.

Since the theory ceases to be logarithmic  in the sub-Ohmic regime, the standard RG approach is no longer directly applicable. 
However, it can be applied to the $s< 1$ case in the spirit of Wilson's $\varepsilon$-expansion. It assumes that a small deviation of the exponent $s$ from unity, $\varepsilon=1-s$, leads to corrections to the critical indices calculated in the case of the ohmic regime. Thus, the Ohmic model is the marginal case, and $\varepsilon$ is treated as a small parameter allowing control of the deviation from marginality.

Taking into account the dimension regularization, we can rewrite the RG equations as follows (see Appendix B): 
\begin{gather}\label{RGEU}
\left\{
\begin{array}{l}
\displaystyle\frac{\mathrm{d}\Delta}{\mathrm{d}\ln\Lambda} \approx \Delta\left(1-\varepsilon-(1+\varepsilon)\frac{\lambda^2C}{4\pi}\right),
\\[10pt]
\displaystyle\frac{\mathrm{d}\lambda}{\mathrm{d}\ln\Lambda} \approx \lambda\left(\frac{1}2-(1+\varepsilon)\frac{\lambda^2C}{4\pi} \right).
\end{array}
\right.  
\end{gather}

Considering the regularization corrections to $\Delta$, linear over $\varepsilon$, estimated near the WF fixed point ($(1+\varepsilon)C{{\lambda}^*}^2/4\pi\approx 1/2$, $\Delta^*\approx 0$), we have the following condition for the WF fixed point:
\begin{gather}
\label{RGE}
\displaystyle\frac{\mathrm{d}\Delta}{\mathrm{d}\ln\Lambda} \approx \Delta\left(\frac12-\varepsilon\right)=0.
\end{gather}
From this, it follows that at $\varepsilon=1/2$ ($s= 1/2$) the WF fixed point becomes marginal. In the phase transition theory, this corresponds to a transition from non-trivial critical behavior to mean-field behavior.
Thus, $s=1/2$ corresponds to the upper critical dimension, and at $s\leq 1/2$ the theory should demonstrate mean-field behavior. 
{ Certainly, the extrapolation to $\varepsilon = 1/2$ (i.e., $s = 1/2$) in the one-loop approximation should be viewed as an estimate rather than an exact result. However, this estimate agrees well with the calculation results \cite{PhysRevLett.108.160401,shen2023numerical}.
}

{ 
Since the WF fixed point in our system is practically unattainable; therefore,  the critical exponents of the system cannot be determined directly from it. 
}

\begin{figure}[ht]
   \centering
   \includegraphics[scale=0.8]{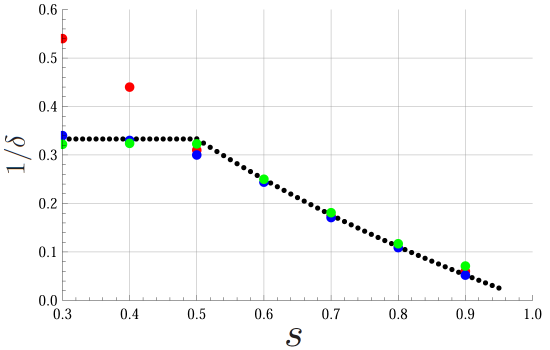}
   \caption{Dependence of the critical exponent $1/\delta$ on $s$: red dotes --- NRG \cite{PhysRevLett.94.070604}, blue dotes --- VMPS \cite{PhysRevLett.108.160401}, green dotes --- NVM\cite{shen2023numerical}, black dotes line is the theoretical results (quantum-classical mapping and our approach).}
   \label{MN}
\end{figure}

However, the critical exponent of magnetization $\delta$ 
can be estimated in the one-loop approximation using the expression for the correlation function of the $\psi^{cl}$-field  $\langle\psi^{cl}_y\psi^{cl}_z\rangle_{\omega}$ (see Appendix A). 
To do this, we assume that the system is close to the critical point, but the observation time is relatively long compared to the coherence time, $\omega<\omega'< \Delta$. In the vicinity of $\lambda=\lambda_c$ the renormalized  $\Delta=\Delta^*\to 0$. Then, on the one hand, 
\begin{gather*}
\langle \sigma_x\rangle=-\mathrm{i}\int\limits^{\omega'}_0\dfr{\mathrm{d}\omega}{2\pi}\langle\psi^{cl}_y\psi^{cl}_z\rangle_{\omega}=
\int\limits^{\omega'}_0\dfr{\mathrm{d}\omega}{2\pi}\dfr{h}{\Delta^2+h^2-\omega^2}\propto \omega'/h.
\end{gather*}
On the other hand, from (\ref{R16}) near the transition between the localized and delocalized states
\begin{gather}
\lambda^2\int\limits_{0}^{\omega'}\dfr{\mathrm{d}\omega}{2\pi}\dfr{C\,|\omega|^{s}}{\Delta^2+h^2-\omega^2}\propto \omega'^{(s+1)}/h^2\sim 1,
\end{gather}
therefore $\omega'\propto h^{{2}/({s+1})}$ and we can conclude that $\langle \sigma_x\rangle \propto h^{{(1-s)}/{(s+1)}}$. This results in the critical exponent of magnetization being
$
1/\delta \approx {(1-s)}/{(s+1)}
$, which is consistent with the results of the numerical calculations of the renormalization group \cite{PhysRevLett.94.070604} and with the numerical calculations using VMPS \cite{PhysRevLett.102.030601,PhysRevLett.108.160401,PhysRevB.81.121105,PhysRevB.100.115106} and NVM \cite{shen2023numerical} (Fig.\,\ref{MN}).
\begin{figure}[ht]
   \centering
   \includegraphics[scale=0.8]{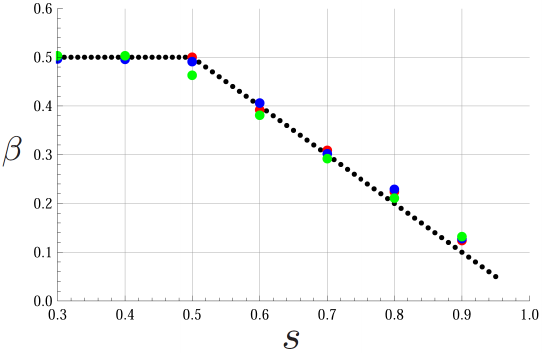}
   \caption{Dependence of the critical exponent $\beta$ on $s$: red dotes --- theoretical values from the quantum-classical mapping \cite{PhysRevLett.29.917,PhysRevLett.108.160401}, blue dotes --- VMPS \cite{PhysRevLett.108.160401}, green dotes --- NVM \cite{shen2023numerical}, black dotes line is our theoretical estimation.}
   \label{MN2}
\end{figure}

{
Another critical exponent of magnetization $\beta$ can be estimated by expanding the right hand side of the second equation in (\ref{RGEU})
near $\lambda_c$. We obtain  ($\delta\lambda=\lambda-\lambda_c$):
\begin{gather}\label{RGEU2}
\displaystyle\frac{\mathrm{d}\,\delta\lambda}{\mathrm{d}\ln\Lambda} \approx \lambda_c\left(\varepsilon-\frac{1}2\right)-\delta\lambda\left(3\varepsilon-\frac{5}2 \right).
\end{gather}
Hence, in the first approximation $|\lambda-\lambda_c|\sim \ln(\Lambda)$, and since $\langle\sigma_x\rangle=-\mathrm{i}\langle\psi_y\psi_z\rangle/2 \sim \ln(\Lambda)^{\varepsilon}$, we can estimate $\langle\sigma_x\rangle\sim |\lambda-\lambda_c|^{1-s}$, i.e. $\beta\approx 1-s$. We arrive at a good result for the first approximation (see Fig.\,\ref{MN2}).   

}

\section{Dissipation}

\begin{figure}[ht]
   \centering  \includegraphics[scale=0.249]{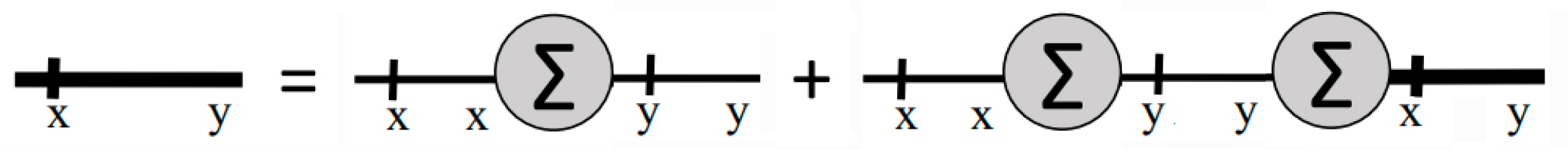}
   \caption{
  Graphical representation of the Dyson equation for the correlation function $\langle\langle\psi_x^q\psi_y^{cl}\rangle\rangle$. The thick graph corresponds to the renormalized (``dressed'') correlation function, the thin one to the non-renormalized one, $\Sigma$ is the self-energy part. }
   \label{Dayson}
\end{figure}

\begin{figure}[ht]
   \centering
   \includegraphics[scale=0.15]{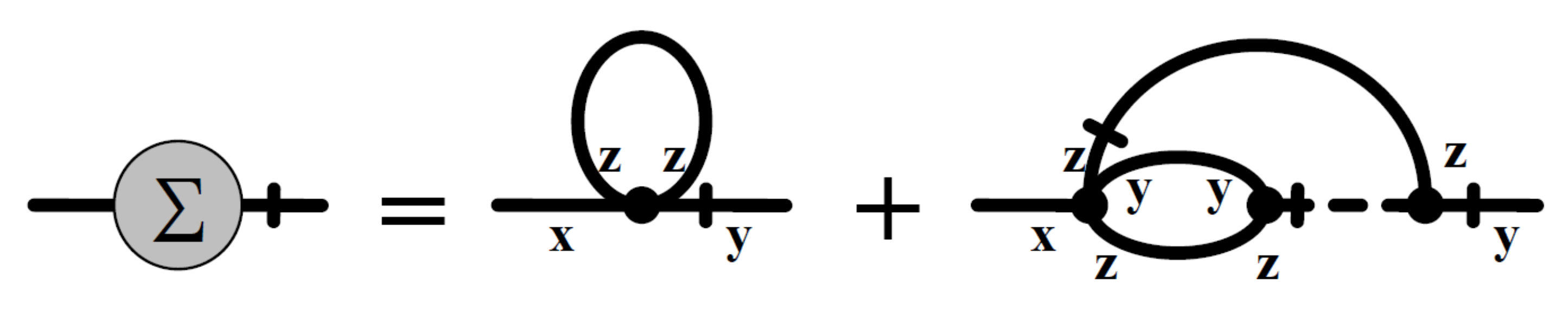}
   \caption{
  Diagrammatic representation of the first contribution to the self-energy $\Sigma$. }
   \label{Defaz2}
\end{figure}

Above, we considered the renormalization $\Delta$ (the Lamb shift) arising from the interaction of spin with the bosonic heat reservoir. Let us now consider the dissipation caused by this interaction. It is known that this dissipation is characterized by two timescales: the decoherence time and the relaxation time. To estimate the former, we consider the renormalization of the correlation function $\langle\langle\psi^q_x\psi^{cl}_y\rangle\rangle$. To do this, we represent the renormalized correlation function using the Dyson equation (see Fig.\,\ref{Dayson}), in which the self-energy term is the sum of two contributions (see Fig.\,\ref{Defaz2}). The first of these is equal to $\Delta$ since $\psi_z\psi_z=1/2$. The second corresponds to the dissipation associated with the interaction of spin with the bosonic heat reservoir:
\begin{gather}\label{F2}
\Sigma=\Delta-\mathrm{i}\Delta\lambda^2C.    
\end{gather}
As a result, the renormalized correlation function is written as follows:  
\begin{gather}\label{F1}
\langle\langle\psi_x^{cl}\psi_y^{q}\rangle\rangle_{\omega}=\frac{\mathrm{i}}2\frac{\Sigma}{(\omega-\Sigma)(\omega+\Sigma)}  .  
\end{gather}

From (\ref{F2}) and (\ref{F1}) one can see: in the region $\lambda<\lambda^*=\sqrt{2\pi/C}$, the dephasing time $1/\Delta\lambda^2C$ is greater than the oscillation period $2\pi/\Delta$, so this region corresponds to the damped coherent state. In the region $\lambda>\lambda^*$, the dephasing time is less than the tunneling time, and the system is in the incoherent state.
This conclusion completely agrees with the results of Leggett's theory, considering that $\lambda=\lambda^*$ corresponds to $\alpha=1/2$ in \cite{RevModPhys.59.1}. 

    {
While the above analysis of dissipation is presented for the ohmic case $s=1$ for simplicity, the Majorana--Keldysh formalism is directly applicable to the sub-ohmic regime $s<1$. In that case, the self-energy $\Sigma$ acquires a non-trivial dependence on $s$. Consequently, the crossover condition between the damped oscillatory and incoherent regimes becomes $s$-dependent, generalizing the ohmic result $\alpha=1/2$.  Based on RG equations (\ref{RGEU}) and the form of the self-energy $\Sigma $, one can expect that at $s\approx 1/2$, when $\lambda^*$ and $\lambda_c$ have coincided, the real and imaginary parts of the self-energy (\ref{F2}) will also be close. Thus, at $s<1/2$, only the oscillatory regime will persist due to the finiteness of the cutoff frequency $\omega_D$ in real solid-state systems~\cite{PhysRevB.81.054308,PhysRevLett.110.010402,PhysRevLett.129.120406}. A detailed analysis of this dynamical crossover in the sub-ohmic regime is beyond the scope of the present work but can be systematically addressed within our framework in the future. 
    }

The relaxation time is the timescale over which the average value of the z-component of the spin, $\sigma_z$, relaxes. Using the Majorana technique we are considering, the expression for this correlation function is represented as (see Appendix D):
\begin{multline}\label{BR}
\langle \sigma_z\sigma_z\rangle_{\omega}=-4\langle\langle {\psi^{cl}_x\psi^{cl}_y\psi^{cl}_z\psi^{cl}_z},\,{\psi^{cl}_x\psi^{cl}_y\psi^{cl}_z\psi^{cl}_z}\rangle\rangle_{\omega}\\
=2\pi\delta(\omega)\langle \sigma_z\rangle^2+
\left(1-4\langle \sigma_z\rangle^2\right)\frac{\mathrm{i}\Sigma}{\omega^2-4\Sigma^2},
\end{multline}
which is the Bloch--Redfield equation, the derivation of which using the Majorana technique was also discussed in \cite{PhysRevB.93.174420}. From this expression, one can see that the relaxation time is twice the dephasing time, which agrees with what is already known.

\section{Discussion and Conclusions}

This study presents an alternative method for analyzing the spin-boson model, based on spin fermionization using Majorana fermions (spinors) and the application of critical dynamics and renormalization group analysis within the Schwinger--Keldysh formalism. First of all, the problem of renormalizability has been overcome. It was shown that the standard Hamiltonian formulation in terms of Majorana fermions leads to a non-renormalizable theory. To solve this problem, an equivalent but renormalizable representation of the Hamiltonian was proposed, which correctly accounts for the commutation relations of the spin operators in the diagrammatic technique. This allowed us, within the one-loop renormalization group approximation, to derive equations for the renormalization of the tunneling constant $\Delta$ and the coupling constant $\lambda$. 
{
 It is important to note that the RG equations are derived within a one-loop (weak-coupling) approximation. However, in the critical region near the quantum phase transition, the renormalization group allows us to sum the leading logarithmic divergences to all orders, effectively extrapolating the weak-coupling results into the strong-coupling regime. In our case, the validity of this approach is supported by the agreement of our analytical predictions for the critical exponents and the critical coupling $\lambda_c(s)$ with numerical data obtained from methods like NRG, DMRG, VMPS and NVM \cite{PhysRevLett.94.070604, PhysRevLett.102.030601,doi:10.1080/14786430500070396,PhysRevLett.108.160401, PhysRevB.81.121105, PhysRevB.85.144425,PhysRevB.100.115106,PhysRevLett.91.170601,shen2023numerical} especially for $s$ close to $1$.
    }

The theory agrees with Leggett's theoretical statements \cite{RevModPhys.59.1} and consistently describes three regimes depending on the spectral function exponent $s$:
\begin{itemize} 
    \item For the Ohmic regime ($s=1$), the identified critical point is $\lambda^*$. It is shown that the transition between localized and delocalized phases occurs at $\lambda=\lambda_c\equiv\sqrt{2}\lambda^*$, which corresponds to $\alpha=1$ in Leggett's theory; 
    \item In the Sub-Ohmic regime ($s<1$), as in the Ohmic regime, there exists a transition between the delocalized phase (for $\lambda < \lambda_c$) and the localized phase (for $\lambda > \lambda_c$).  We show that $\lambda_c$ exponentially decreases as $s$ decreases, narrowing the delocalization region. The renormalization group analysis leads to the conclusion that at $\varepsilon=1/2$ ($s= 1/2$) the WF fixed point becomes marginal. In phase transition theory, this corresponds to a transition from non-trivial critical behavior to mean-field behavior. Thus, $s=1/2$ corresponds to the upper critical dimension, and at $s\leq 1/2$, the model should demonstrate mean-field behavior.  Aside from that, the expression for the critical magnetization exponent derived within this framework has the hyperscaling form: $1/\delta = (1-s)/(1+s)$. When $s$ reaches the value $s=1/2$, the critical exponent $\delta$ also coincides with its mean-field value. All these agree with the quantum-classical mapping and with the results of numerical calculations for $s\lesssim 1$ \cite{PhysRevLett.94.070604, PhysRevLett.102.030601,doi:10.1080/14786430500070396,PhysRevLett.108.160401, PhysRevB.81.121105, PhysRevB.85.144425,PhysRevB.100.115106,PhysRevLett.91.170601,shen2023numerical}.
    \item  In the Super-Ohmic regime ($s>1$), the localized phase is absent, and the spin always remains delocalized, as the parameters $\Delta$, $h$, and $\lambda$ are not renormalized.
\end{itemize}

In terms of the presented approach, the Bloch--Redfield equation was explicitly derived, and it was shown that the relaxation time is twice the decoherence time, which also agrees with accepted notions.
 {
Note that although the Bloch--Redfield equation is a standard result in the theory of open quantum systems, its derivation within the Majorana--Keldysh framework is not merely a reformulation. It emerges naturally from the same renormalization procedure used to analyze the critical properties, thereby establishing a unified description of equilibrium criticality and nonequilibrium dissipation.
}

In addition, the proposed approach provides a comprehensive description of dissipation processes, allowing for the calculation of relaxation and decoherence times. It is shown that:
\begin{itemize}
\item
For $\lambda<\lambda^*$ ($\alpha<1/2$ in \cite{RevModPhys.59.1}), the decoherence time is longer than the period of tunneling oscillations, corresponding to a damped oscillatory regime;
\item For $\lambda>\lambda^*$ ($\alpha>1/2$ in \cite{RevModPhys.59.1}), the decoherence time is shorter than the tunneling time, corresponding to an incoherent relaxation regime.
\end{itemize}
This is also in good agreement with Leggett's theory \cite{RevModPhys.59.1}.

Our results generally agree with those of the theory based on quantum-classical mapping but differ in the determination of the type of phase transition. According to quantum-classical mapping, the phase transition between the localized and delocalized phases occurs directly at $\alpha =1$ and represents a Berezinskii--Kosterlitz--Thouless transition. We also consider the model's criticality close to the Ohmic regime, where we can use Wilson's UV-renormalization group method. The identified WF fixed point $\lambda=\lambda^*$ corresponds to a second order phase transition. However, the transition between the delocalized and localized states does not happen in this point, but in the point $\lambda_c>\lambda^*$ in which the infrared behavior of the RG flow is violated as a result of the quantization of the spin-boson interaction.

This discrepancy likely stems from the differences in the approximations inherent to each method. On the one hand, our approach uses the adiabatic approximation, which violates the infrared behavior of the RG flow. On the other hand, the quantum-classical mapping, which captures the long-range interactions, may not fully account for the short-range transverse field terms, which could become relevant in the extreme quantum limit \cite{PhysRevB.9.215, PhysRevLett.94.070604}. In both cases, approximations are used that differ from one another but lead to similar results.

Thus, the presented formalism, which marries the Majorana representation with techniques of critical dynamics, constitutes a consistent theoretical framework. It enables a unified analysis of criticality and dissipation in the spin-boson model across regimes adjacent to the ohmic point ($s \lesssim 1$), circumventing the need for quantum-classical mapping. The key analytical predictions of this framework align well with findings from alternative computational and theoretical approaches.

Future work could extend this approach to multi-spin systems, non-equilibrium dynamics, finite-temperature effects, and systems with structured spectral densities. The Majorana representation, combined with the Schwinger-Keldysh formalism, provides a powerful tool for studying quantum impurity systems, offering new insights into the interplay between criticality and dissipation.

\section*{APPENDIX A: Interaction part of the action}

The system's action is
\begin{equation}
\mathcal{S}=\frac{\mathrm{i}}{2}\int \mathrm{d}t\,\vec\psi\partial_t\vec\psi+\mathcal{S}_{int}+\mathcal{S}_{bath}, 
\label{action0}
\end{equation}
where $\mathcal{S}_{bath}$ is the action part corresponding to the bosonic bath, and $\mathcal{S}_{int}$ is the action part that includes the environmental coupling to the spin.

Combining terms quadratic in $\psi$, one can write expressions for the initial correlation functions of the model, ${\bf G}(t)$, which take into account the part of the action that is disconnected from the bosonic bath.
Within the Matsubara formalism, using the Hamiltonian (\ref{Ham}), the action can be written as
\begin{equation}
    \mathcal{S}=-\frac{1}{2}\int_0^{\beta}d\tau\left[ \vec\psi\:{\bf G}^{-1}\vec\psi+\mathrm{i}\lambda X\psi_y\psi_z\right]+\mathcal{S}_{bath},
\end{equation}
where
\begin{gather*}
{\bf G}^{-1}
=\left(\begin{array}{ccc}
    -\partial_{\tau} & -\mathrm{i}\Delta & 0\\[12pt]
    \mathrm{i}\Delta &  -\partial_{\tau} & -\mathrm{i}h \\[12pt]
    0 & \mathrm{i}h &  -\partial_{\tau}
\end{array}\right).
\end{gather*}
In frequency representation, we get
\begin{equation}
{\bf G}^{-1}(\omega_n)
=\left(\begin{array}{ccc}
    \mathrm{i}\omega_n & -\mathrm{i}\Delta & 0\\[12pt]
    \mathrm{i}\Delta &  \mathrm{i}\omega_n  & -\mathrm{i}h \\[12pt]
    0 & \mathrm{i}h &  \mathrm{i}\omega_n 
\end{array}\right).
\end{equation}
Inverting this matrix, we obtain
\begin{multline*}
{\bf G}(\omega_n)
\\=-\frac{1}{2\omega_n}\frac{\mathrm{i}}{\Delta^2+h^2+\omega_n^2}\left(
\begin{array}{ccc}
    h^2+\omega_n^2 & \Delta\omega_n & h\Delta\\[12pt]
    -\Delta\omega_n & \omega_n^2 & h\omega_n \\[12pt]
    h\Delta & -h\omega_n & \,\Delta^2+\omega_n^2
\end{array}
\right),
\end{multline*}
Making an analytical continuation to the real axis $\mathrm{i}\omega_n\to \omega\pm \mathrm{i}\delta$, we arrive at the result
\begin{multline*}
{\bf G}(\omega)
\\=\frac{1}{2\omega}\frac{1}{\Delta^2+h^2-\omega^2}\left(
\begin{array}{ccc}
    h^2-\omega^2 & -\mathrm{i}\Delta\omega &h\Delta\\[12pt]
    \mathrm{i}\Delta\omega & -\omega^2 &-\mathrm{i}h\omega \\[12pt]
    h\Delta & \mathrm{i}h\omega & \,\,\,\Delta^2-\omega^2
\end{array}
\right).
\end{multline*}

We consider the bosonic bath as a system of harmonic oscillators described by the scalar field $X$. After the "Wick rotation", the bosonic part of the action has the following form:
\begin{equation} 
\mathcal{S}_{bath}= \frac{1}{2}\sum_i\int\mathrm{d} t\, X_i D_iX_i,
\end{equation}
where the bath oscillators are labeled by an index $i$, and
\begin{equation}
D_i=\partial_t-\mathrm{i}\omega_i,
\end{equation}
with $\omega_i$ being the energy of the $i$-th oscillator.
The bosonic correlation function of this system has the following form:    
\begin{equation}
\langle XX\rangle_{\omega}=\frac{1}{2\mathrm{i}}\sum_i\frac{1}{\omega+ \mathrm{i}\epsilon-\omega_i}=\int^{\infty}_{-\infty}\frac{\mathrm{d}x}{2\pi\mathrm{i}}\frac{f(x)}{\omega+ \mathrm{i}\epsilon-x},   
\label{XX}
\end{equation}
where $\epsilon \to 0$, and $f(\omega)=\sum_i\delta(\omega-\omega_i)$ is the bath spectral function.
The case of $f(\omega)\propto \omega$ corresponds to the so-called ohmic case, which is frequently found in more realistic models of the environment. In the general case, the bath spectral density can be expressed as $f(\omega)\propto\omega^s$; thus, 
$\langle XX\rangle_{\omega}=\mathrm{i}C\,\omega^s$, where $C$ is a dimensionless constant parameter. {
It is related to the standard dissipation strength $\alpha$ used in the spin-boson literature by $C = 2\alpha / (\pi \lambda^2\omega_D^{1-s})$, where $\omega_D$ is the Debye cutoff frequency.
}

We investigate the critical behavior of the presented model in terms of the Schwinger--Keldysh technique \cite{Kamenev_2011}, denoting by symbols $\psi^+$ and $\psi^-$ the Majorana spinor field on the forward and backward branches of the Keldysh contour. By performing a Keldysh rotation, we rewrite the model in new fields 
$\psi^{cl}=(\psi^++\psi^-)/\sqrt{2}$, $\psi^{q}=(\psi^+-\psi^-)/\sqrt{2}$, $X^{cl}=(X^++X^-)/\sqrt{2}$, $X^{q}=(X^+-X^-)/\sqrt{2}$,
and represent the system's action as follows:
\begin{equation}
\mathcal{S}^{SK}=\frac{\mathrm{i}}{2}\int \mathrm{d}t\,(\vec\psi^{q}\partial_t\vec\psi^{cl}- \vec\psi^{cl}\partial_t\vec\psi^{q})+\mathcal{S}^{SK}_{int}+\mathcal{S}^{SK}_{bath}, 
\label{action}
\end{equation}
where 
\begin{multline}\label{oldS}
\mathcal{S}^{SK}_{int}=\frac{\mathrm{i}}{2}\int \mathrm{d}t\,\left[
{\mathrm{i}\Delta}\left({\psi}^{q}_x\psi_y^{cl}+{\psi}^{cl}_x\psi_y^{q}\right)\right.\\+
{\mathrm{i}(\lambda X^{cl}}+h)\left(\psi_y^{q}\psi_z^{cl}+\psi_y^{cl}\psi_z^{q}\right)\\+
\left.\mathrm{i}(\lambda X^{q}+h)\left({\psi}^{cl}_y\psi_z^{cl}+\psi_y^{q}\psi_z^{q}
\right) \right].
\end{multline}
The bosonic part of the action, after the ``Wick rotation'', has the following form:
\begin{equation} 
\mathcal{S}^{SK}_{bath}= \frac{1}{2}\sum_i\int\mathrm{d} t\, \left(X^q_i D_iX^{cl}_i-X^{cl}_i {D_i^*}X^{q}_i\right).
\end{equation}
The retarded $\langle XX\rangle^{R}\equiv\langle X^{q}X^{cl}\rangle$ and advanced $\langle XX\rangle^{A}\equiv\langle X^{cl}X^{q}\rangle$ correlation functions of this system have the following form:    
\begin{equation}
\langle XX\rangle^{R/A}_{\omega}=\int^{\infty}_{-\infty}\frac{\mathrm{d}x}{2\pi\mathrm{i}}\frac{f(x)}{\omega\pm \mathrm{i}\epsilon-x},   
\label{XX}
\end{equation}
thus 
$\langle XX\rangle^{R/A}_{\omega}=\pm\mathrm{i}C\,\omega^s$.  Since the boson bath is in equilibrium, the Keldysh component of the correlator is determined by the fluctuation-dissipation theorem:
\begin{equation*}
\langle XX\rangle_{\omega}^K=\left(\langle XX\rangle^R_{\omega}-\langle XX\rangle^A_{\omega}\right)\coth \frac{\omega}{2T}=2\mathrm{i}C\,\omega^s\coth \frac{\omega}{2T}.
\end{equation*}

It is not difficult to check that $\{\psi^{\alpha}_i\psi_j^{\beta}\}= \delta_{ij}$, $[\psi^{\alpha}_i\psi_j^{\beta}]=2(1-\delta_{ij})\psi^{\alpha}_i\psi^{\beta}_j$, where $\alpha,\beta=(cl,q)$.
For future convenience, we use a Euclidean formulation, which can be obtained by "Wick rotation" $t \to \mathrm{i}t$, and we write the bare retarded and advanced causal propagators of the massless Majorana spinor field as
\begin{equation*}
\langle\psi^{cl}_i\psi^{q}_j\rangle^0_{\omega}=\frac{1}{2}\frac{\delta_{ij}}{\epsilon+\mathrm{i}\omega},\quad \langle\psi^{q}_i\psi^{cl}_j\rangle^0_{\omega}=\frac{1}{2}\frac{\delta_{ij}}{\epsilon-\mathrm{i}\omega}.
\end{equation*}

It is important to note that the Majorana fermions are auxiliary fields that are not thermalized on their own terms. They thermalize only by pairs, forming the spin that is being thermalized. Therefore, the causal propagators of the Majorana fermion subsystem are assumed to correspond to zero temperature, $T_{M}=0$. Thus, 
\begin{multline*}
\langle\psi^{cl}_i\psi^{cl}_j\rangle^0_{\omega}=\left[\langle\psi^{cl}_i\psi^{q}_j\rangle^0_{\omega}-\langle\psi^{q}_i\psi^{cl}_j\rangle^0_{\omega}\right]\tanh(\omega/2T_{M})\\=
-\frac{\mathrm{i}\omega\delta_{ij}}{\omega^2+\epsilon^2}, \quad \langle\psi^{q}_i\psi^{q}_j\rangle^0_{\omega}=0 \quad (i,j=x,\,y,\,z).
\end{multline*}

\begin{figure}[ht]
   \centering
   \includegraphics[scale=0.45]{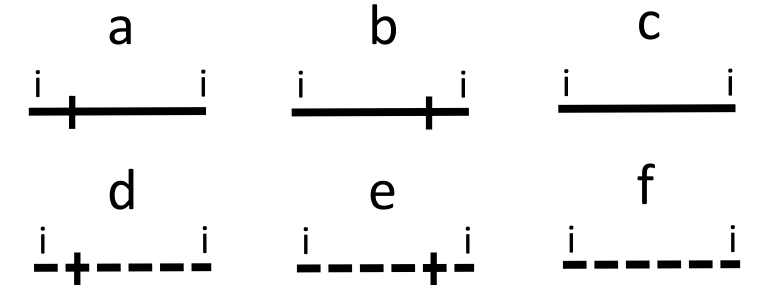}
   \caption{Graphical presentation of the propagators: a) $\langle\psi^{q}_i\psi^{cl}_i\rangle^0$; b) $\langle\psi^{cl}_i\psi^{q}_i\rangle^0$; c) $\langle\psi^{cl}_i\psi^{cl}_i\rangle^0$; d) $\langle X^{q}X^{cl}\rangle$; e) $\langle X^{cl}X^{q}\rangle$; f) $\langle X^{cl}X^{cl}\rangle$.}
   \label{F}
\end{figure}

\begin{figure}[ht]
   \centering
   \includegraphics[scale=0.27]{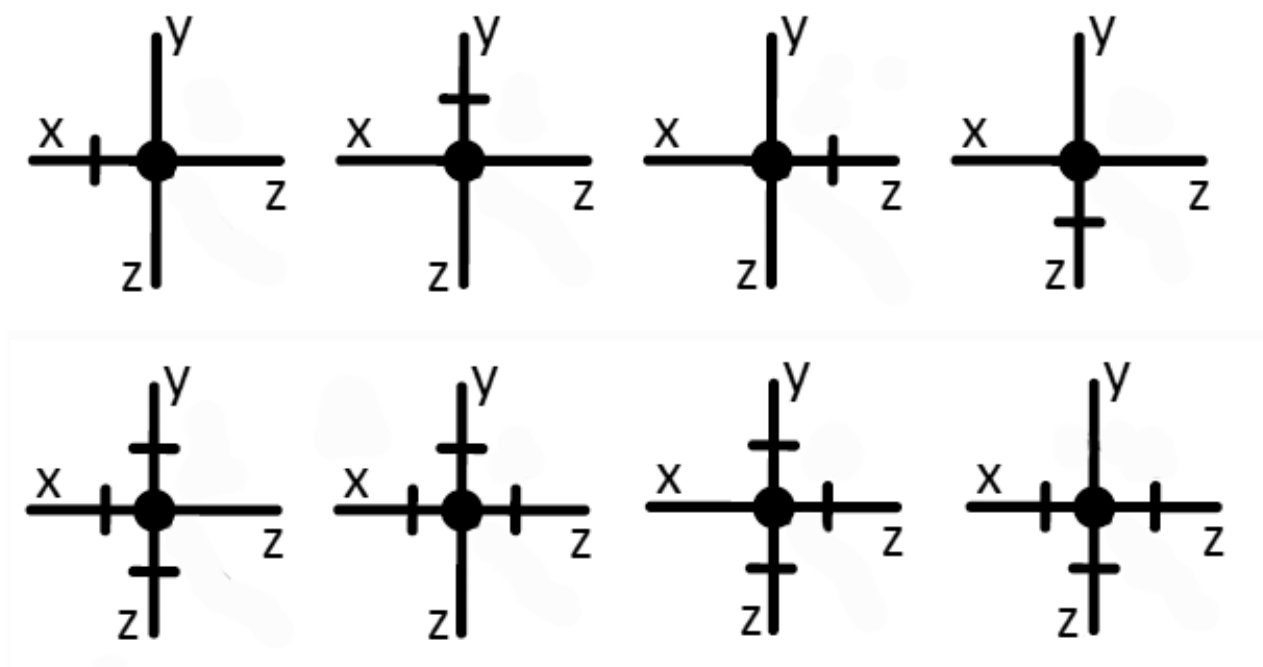}
   \caption{Graphical presentation of the $\Delta$ vertices. The line with the stroke corresponds to the "quantum" field $\psi^q$.}
   \label{F01}
\end{figure}

\begin{figure}[ht]
   \centering
   \includegraphics[scale=0.4]{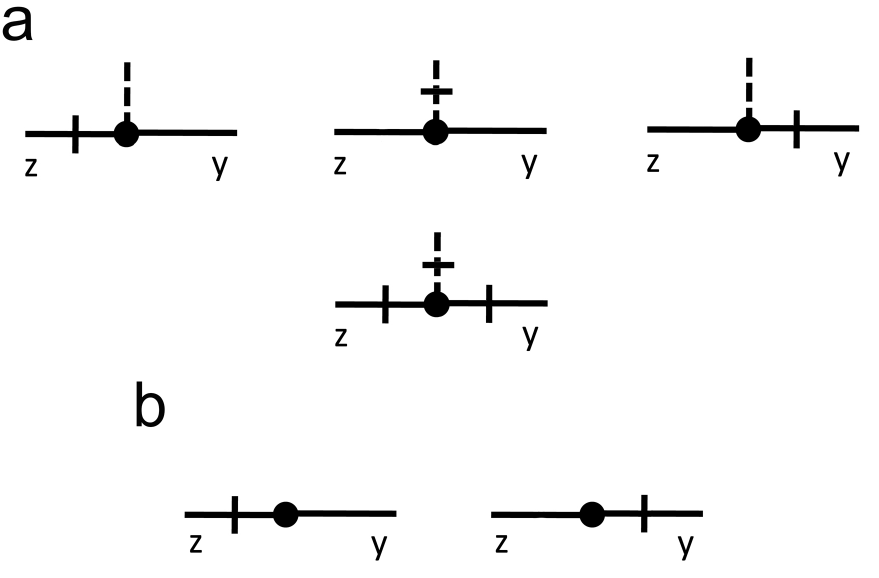}
   \caption{Graphical presentation of the $\lambda$ (a) and $h$ (b) vertices. The lines with the stroke correspond to the "quantum" fields ($\psi^q$ or $X^q$).}
   \label{1F1}
\end{figure}

The theory contains two types of propagators (Fig.~\ref{F}) and three types of vertices: the vertices $\Delta$ (Fig.~\ref{F01}); and $\lambda$ (Fig.~\ref{1F1}).  

Combining the quadratic terms and using the Dyson equation, we can write expressions for the initial correlation functions, which do not take into account the influence of the bosonic bath: 
\begin{gather*}
\langle\psi^{cl}\psi^{q}\rangle_{\omega}=
{\langle\psi^{q}\psi^{cl}\rangle_{\omega}^*}
={\bf G}(\omega).
\end{gather*}

The interaction from the bosonic heat bath leads to a renormalization of the vertices, which in the one-loop approximation looks like this (see Fig.~\ref{PF2}):
\begin{align}
{\Delta}' &\approx \Delta-\frac{\Delta\lambda^2}{2}\int_{\omega'}^{\omega_D}\frac{\mathrm{d}\omega}{2\pi}\frac{C\omega^{s}(\Delta^2-\omega^2)}{(\Delta^2+h^2-\omega^2)^2},\\
{h}' &\approx h-\frac{h\lambda^2}{2}\int_{\omega'}^{\omega_D}\frac{\mathrm{d}\omega}{2\pi}\frac{C\omega^s}{\Delta^2+h^2-\omega^2}, \\
{\lambda}' &\approx \lambda -\frac{\lambda^3}{2}\int_{\omega'}^{\omega_D}\frac{\mathrm{d}\omega}{2\pi}\frac{C\omega^{s}(\Delta^2-\omega^2)}{(\Delta^2+h^2-\omega^2)^2}, 
\end{align}
where $\omega_D$ is the upper cutoff limit (Debye frequency) and $\omega'$ is the lower limit of integration.
We assume that the system undergoes a phase transition between localized and delocalized states, and we consider the fluctuation regime of this transition below.

\begin{figure}[ht]
   \centering
   \includegraphics[scale=0.45]{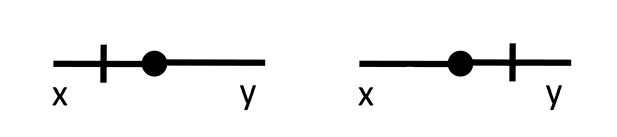}
   \caption{Graphical presentation of the $\Delta$ vertices corresponding to (\ref{oldS}). The line with the stroke corresponds to the "quantum" field $\psi^q$.}
   \label{F0}
\end{figure}

We begin by examining the case in which the system Hamiltonian takes the form (\ref{Ham}). In this case, in the Keldysh technique, the interaction part of the system's action has the form presented in (\ref{oldS}).
The diagrammatic representation of the causal propagators for Majorana fermions and bosons is shown in Fig.~\ref{F}, and the diagrammatic representation of the $\Delta$, $h$ and $\lambda$ vertices is shown in Figures \ref{F0} and \ref{1F1}, respectively.

\begin{figure}[ht]
   \centering
   \includegraphics[scale=0.37]{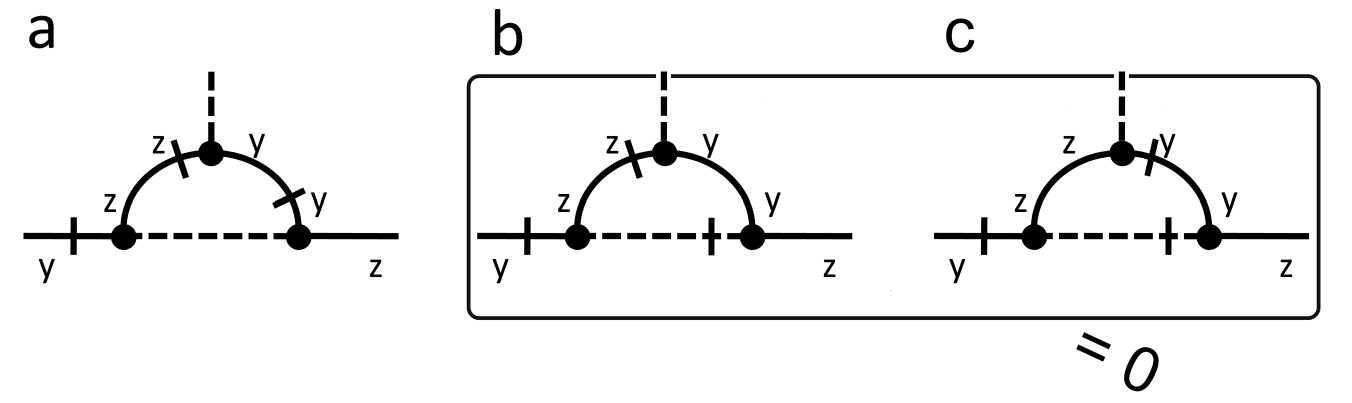}
   \caption{Diagrams of first level contributions to the renormalization of $\mathrm{i}\lambda\psi^{q}_y\psi^{cl}_z$ vertex. The contributions from diagrams (b) and (c) have opposite signs, resulting in their mutual cancellation.}
   \label{FA0}
\end{figure}

One can see that the $\lambda$ vertex is being renormalized. In the one-loop approximation, the renormalization term has the diagrammatic form presented in Fig.~\ref{FA0}.  
The $h$ is also being renormalized; its one-loop renormalization term is presented in Fig.~\ref{PF2}\,(b).

However, it is obvious that this set of propagators and vertices does not allow for the construction of a diagram that renormalizes the vertex $\Delta$. Thus, the theory described by this form of Hamiltonian is non-renormalizable. 
This may seem strange since the described system is quite simple, and the theory with this Hamiltonian well describes all its properties. 

We think that, in our case, the reason for the non-renormalizability of the theory lies in the fact that the Hamiltonian itself does not contain information about the commutation properties of the field $\psi$, and accounting for these properties is critically necessary for correct renormalization. A deeper understanding of the mathematical aspect of this issue requires additional research. In our case, however, we will limit ourselves to preserving physical accuracy and representing the vertex $\Delta$ as a product of four spinors:
\begin{equation}
\Delta \sigma_z=\Delta\Theta\psi_z=-\mathrm{i}\Delta\, 2\psi_x\psi_y\psi_z\psi_z.
\end{equation}
It can be seen that this representation is equivalent to the previous one since $\psi_z\psi_z=1/2$. At the same time, it contains information about the commutation relations of the spin components:
\begin{multline}
\sigma_x =-2\mathrm{i}\,\psi_x\psi_y\psi_z\psi_x\\
=-\mathrm{i}\,\psi_x\psi_y\psi_z\psi_x+\mathrm{i}\,\psi_z\psi_x\psi_x\psi_y\\
=\mathrm{i}\left(\sigma_z\sigma_y-\sigma_y\sigma_z\right)=-\mathrm{i}[\sigma_y, \sigma_z].
\end{multline}

\begin{figure}[ht]
   \centering
   \includegraphics[scale=0.37]{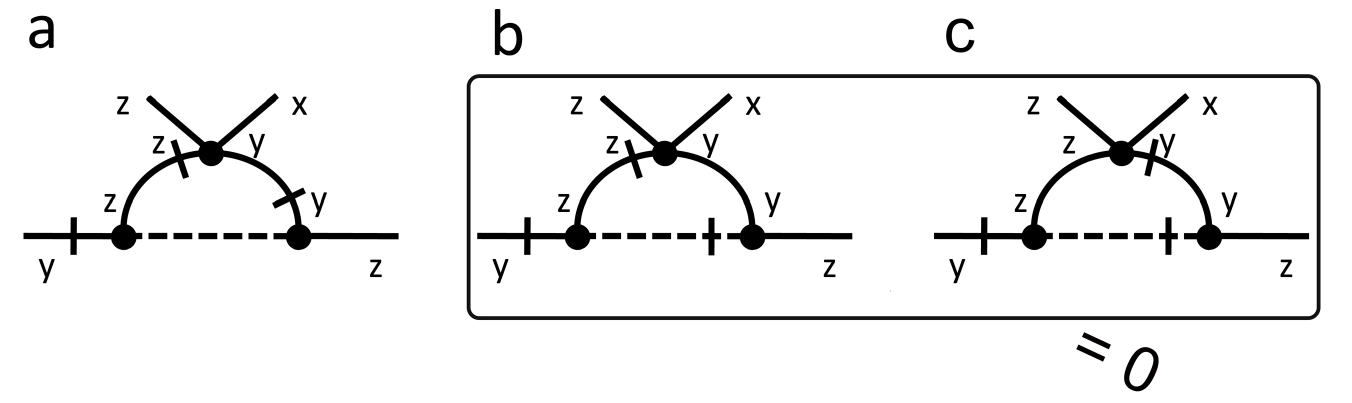}
   \caption{Diagrams of first level contributions to the renormalization of $\mathrm{i}2\Delta\psi^{cl}_x\psi^{q}_y\psi^{cl}_x\psi^{cl}_z$ vertex. The contributions from diagrams (b) and (c) have opposite signs, resulting in their mutual cancellation.}
   \label{FA1}
\end{figure}

Taking into account the above, we rewrite the Hamiltonian (\ref{Ham}) in the following form:
\begin{equation}
\mathcal{H}=
\mathrm{i}2\Delta\psi_x\psi_y\psi_z\psi_z+\mathrm{i}\left(\lambda X+h\right)\psi_y\psi_z+\mathcal{H}_{X}. 
\end{equation}
As a result, the interaction part of the system's action has the following form:
\begin{multline}
\mathcal{S}^{SK}_{int}=\frac{\mathrm{i}}{2}\int \mathrm{d}t\,\left[
{\mathrm{i}2\Delta}\left({\psi}^{q}_x\psi_y^{cl}{\psi}^{cl}_z\psi_z^{cl}+{\psi}^{cl}_x\psi_y^{q}{\psi}^{cl}_z\psi_z^{cl}\right.\right.\\\left.
+2{\psi}^{cl}_x\psi_y^{cl}{\psi}^{q}_z\psi_z^{cl}\right)+
{\mathrm{i}2\Delta}\left({\psi}^{cl}_x\psi_y^{q}{\psi}^{q}_z\psi_z^{q}+{\psi}^{q}_x\psi_y^{cl}{\psi}^{q}_z\psi_z^{q}\right.\\\left.+2{\psi}^{q}_x\psi_y^{q}{\psi}^{q}_z\psi_z^{cl}\right)+\mathrm{i}\lambda X^{q}\left({\psi}^{cl}_y\psi_z^{cl}+\psi_y^{q}\psi_z^{q}
\right)\\+{\mathrm{i}\left(\lambda X^{cl}+h\right)}\left(\psi_y^{cl}\psi_z^{q}+\psi_y^{q}\psi_z^{cl}\right)
\left. \right].
\label{I4}
\end{multline}
In this case, the diagrammatic representation of the $\Delta$ vertex is shown in Figures \ref{F01}.
Now, the theory becomes renormalizable. Figure \ref{FA1} shows the diagrams contributing to the renormalization of the $\Delta$ vertex (exactly $\mathrm{i}2\Delta\psi^{cl}_x\psi^{q}_y\psi^{cl}_x\psi^{cl}_z$). The contributions of two of them cancel each other out, as they are equal in magnitude but have opposite signs. The renormalization of the $\lambda$ vertex remains unchanged.

\section*{APPENDIX B: Renormalization procedure}

Here, we consider the renormalization procedure for the model under consideration and present a renormalization group for its general case, characterized by an arbitrary exponent $s$. 

First of all, using the renormalization equations (\ref{R16}), we can write the renormalization constants of our theory in the following way:
\begin{gather}
Z_{\Delta} \approx \Delta-\frac{\Delta\lambda^2}{2}\int\limits_{\omega'}^{\Lambda\omega'}\dfr{\mathrm{d}\omega}{2\pi}\dfr{C\omega^{s}(\omega^2-\Delta^2)}{(\Delta^2+h^2-\omega^2)^2},\nonumber \\ \label{R1}
Z_{h} \approx h-\frac{h\lambda^2}{2}\int\limits_{\omega'}^{\Lambda\omega'}\dfr{\mathrm{d}\omega}{2\pi}\dfr{C\omega^s}{\Delta^2+h^2-\omega^2},\\\nonumber
Z_{\lambda}\approx \lambda -\frac{\lambda^3}{2}\int\limits_{\omega'}^{\Lambda\omega'}\dfr{\mathrm{d}\omega}{2\pi}\dfr{C\omega^{s}(\omega^2-\Delta^2)}{(\Delta^2+h^2-\omega^2)^2}.
\end{gather}
where  $1<\Lambda\to 1$, 
and $\omega'$ is a cutoff frequency. For simplicity, it is convenient to rewrite these expressions using the causal propagators ($\omega'\gg\Delta^2+h^2$) and to assume $h\to 0$:
\begin{gather}
Z_{\Delta} \approx \Delta-\frac{\Delta\lambda^2}{2}\int\limits_{\omega'}^{\Lambda\omega'}\dfr{\mathrm{d}\omega}{2\pi}\dfr{C\omega^{s}}{\omega^2},\nonumber \\\nonumber
Z_{\lambda}\approx \lambda -\frac{\lambda^3}{2}\int\limits_{\omega'}^{\Lambda\omega'}\dfr{\mathrm{d}\omega}{2\pi}\dfr{C\omega^{s}}{\omega^2}.
\end{gather}

In the Ohmic regime, the spin-boson model is logarithmic, meaning that renormalization corrections do not depend on scale, and its critical behavior can be described using the renormalization group (RG) method.
In the super-Ohmic and sub-Ohmic cases, the theory is not logarithmic, as the renormalization equations depend on the cutoff frequency. However, for $s \approx 1$, one can use dimensional regularization, 
formally assuming that the dimension of the frequency integral is equal to $d=1+\varepsilon$ ($\varepsilon=1-s$):
\begin{gather}
\int\limits_{1}^{\Lambda}\dfr{\mathrm{d} \omega}{2\pi}\dfr{\omega^{s}}{\omega^2}\to\frac12\int\limits_{1}^{\Lambda}\dfr{\mathrm{d}^{1+\varepsilon}\boldsymbol{\omega}}{(2\pi)^{1-\varepsilon}}\dfr{\omega^{s}}{\omega^2}.
\end{gather}
This trick reduces the theory to the logarithmic one. The parameter $\varepsilon$ regularizes the integral. The Ohmic model is the marginal one, and $\varepsilon$ is a small parameter that allows control of the deviation from marginality.

For the regularized case, the canonical dimensions, $[F]_0$, of any fields and parameters, $F$, are determined from the condition of a dimensionless action. They are presented in the following table:
\begin{center}
\begin{tabular}{ |c||c|c|c|c|c|c|c| } 
 \hline
 $\quad F\quad$ & t & $\,\omega,\,\partial_t\,$  & $\quad\psi\quad$ & \,$h$\, & $\quad \Delta\quad$ & $\quad\lambda\quad$ & $\quad X\quad$\\ 
 \hline
 $[F]_0$ & -1 & $1$ & $\varepsilon/2$ & 1 & $1-\varepsilon$ & $1/2$ & $1/2$\\ 
  \hline
\end{tabular}
\end{center}
Here we should note that the naive dimension (classical dimension) of the bosonic field, $[X]_0=s/2$, is determined from the spectral dependence of the correlation function $\langle X_xX_x\rangle_{\omega}\propto \omega^s$ (see (\ref{XX})). However, we assume $d=1+\varepsilon$; therefore, 
\begin{gather}
\langle XX\rangle
^{K}_{\omega}=\frac12\int\limits^{\infty}_{-\infty}\frac{\mathrm{d}^{1+\varepsilon}\boldsymbol{x}}{2\pi\mathrm{i}}\frac{Cx^{s}}{\omega\pm \mathrm{i}\epsilon-x}\propto \omega.   \end{gather}
Thus, in this case, the canonical dimension of the $X$ field does not depend on $s$ and is $[X]=1/2$.

In the fluctuation regime near the critical point of the phase transition, the aforementioned vertices are renormalized. The graphical form of the one-loop contributions to the renormalization is presented in Fig,\,\ref{PF2}. In the case $d=1-\varepsilon$ the renormalization constants are written in the following way:
\begin{gather}
Z_{\Delta} \approx \Delta-\frac{\Delta\lambda^2}{4}\int\limits_{\omega'}^{\Lambda\omega'}\dfr{\mathrm{d}^{1+\varepsilon}\boldsymbol{\omega}}{(2\pi)^{1+\varepsilon}}\dfr{C\omega^{s}}{\omega^2},\nonumber\\ \nonumber
Z_{\lambda}\approx \lambda -\frac{\lambda^3}{4}\int\limits_{\omega'}^{\Lambda\omega'}\dfr{\mathrm{d}^{1+\varepsilon}\boldsymbol{\omega}}{(2\pi)^{1+\varepsilon}}\dfr{C\omega^{s}}{\omega^2}.
\end{gather}
where  $1<\Lambda\to 1$, 
and $\omega'$ is the cutoff frequency.
The integrals in the above expressions can be approximately ($s\approx 1$) calculated as follows: 
\begin{multline}
I=\frac14\int\limits_{\omega'}^{\Lambda\omega'}\dfr{\mathrm{d}^{1+\varepsilon}\boldsymbol{\omega}}{(2\pi)^{1+\varepsilon}}\dfr{\omega^{s}}{\omega^2}
=\frac{\Omega_{1+\varepsilon}}4\int\limits_{\omega'}^{\Lambda\omega'}\dfr{\mathrm{d}\omega}{(2\pi)^{1+\varepsilon}}\dfr{\omega^{s+\varepsilon}}{\omega^2}
\\
\approx\frac{1+\varepsilon}2\int\limits_{\omega'}^{\Lambda\omega'}\dfr{\mathrm{d}\omega}{2\pi}\dfr{1}{\omega}=
\frac1{4\pi}(1+\varepsilon)\ln\Lambda,
\end{multline}
where $\Omega_{n}$ is the $n$-sphere area.
As a result, we write the renormalized vertices as follows:
\begin{multline}\label{AR2}
\Delta^{R} = \Lambda^{1
-\varepsilon} Z_{\Delta} \approx \Lambda^{1
-\varepsilon}\left[\Delta-(1+\varepsilon)\dfr{\Delta\lambda^2C}{4\pi}\ln\Lambda\right]\\\approx
(1+(1-\varepsilon)\ln \Lambda)\left[\Delta-(1+\varepsilon)\dfr{\Delta\lambda^2C}{4\pi}\ln\Lambda\right],
\end{multline}
\begin{multline}
\lambda^{R}=\Lambda^{1/2} Z_{\lambda}  \\\approx
\Lambda^{1/2}\left[\lambda -(1
+\varepsilon)\dfr{\lambda^3C}{4\pi}\ln\Lambda\right]
 \\ 
 =\left(1+\dfr12\ln \Lambda\right)\left[\lambda -(1
+\varepsilon)\dfr{\lambda^3C}{4\pi}\ln\Lambda\right] ,
\end{multline}
where $Z_{F}$ is the renormalization constant of $F$,
from which one can derive the following renormalization group equations:
\begin{gather}\label{AI9}
\left\{
\begin{array}{l}
\displaystyle\frac{\mathrm{d}\Delta}{\mathrm{d}\ln\Lambda} \approx \Delta\left(1-\varepsilon-(1+\varepsilon)\frac{\lambda^2C}{4\pi}\right),
\\[10pt]
\displaystyle\frac{\mathrm{d}\lambda}{\mathrm{d}\ln\Lambda} \approx \lambda\left(\frac{1}2-(1+\varepsilon)\frac{\lambda^2C}{4\pi} \right).
\end{array}
\right.
\end{gather}

\begin{figure}[ht]
   \centering
   \includegraphics[scale=.3]{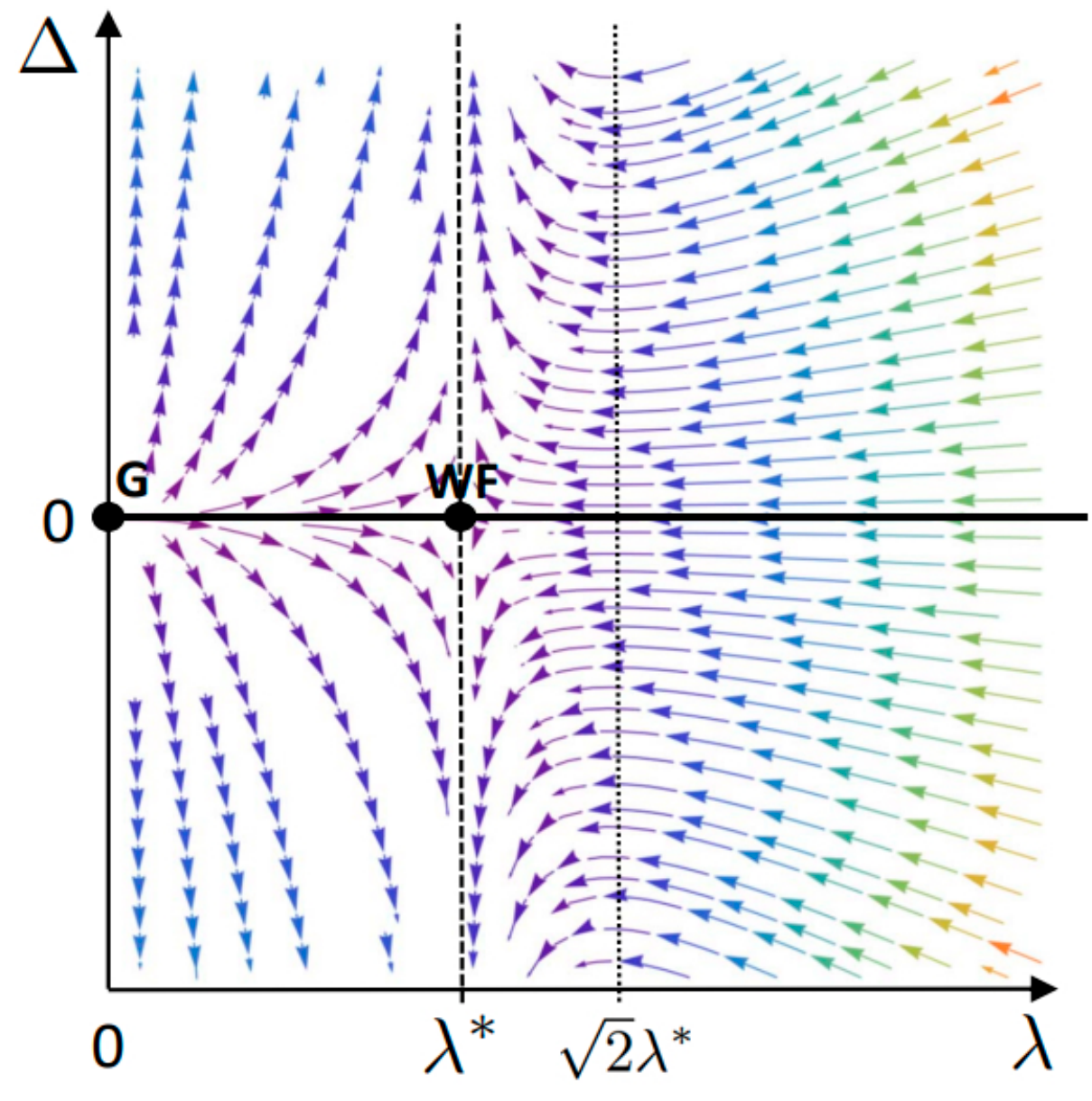}
   \caption{
  Schematic representation of renormalization group (\ref{RG}) flows in the Ohmic case. The critical value $\lambda^*=\sqrt{2\pi/C}$ corresponds to the $\alpha=1/2$ in the Leggett's theory \cite{RevModPhys.59.1}. The extrema of the flow trajectories lie on the line $\lambda=\lambda_c\equiv\sqrt{2}\lambda^*$.}
   \label{S}
\end{figure}

As we noted above, in the Ohmic case ($s=1$), the theory is logarithmic. In this case, the infrared singularities coincide with the ultraviolet singularities, allowing us to use the RG method to analyze the critical behavior of the system.  
The corresponding system of RG equations is written as follows:
\begin{gather}\label{RG}
\left\{
\begin{array}{l}
\displaystyle\frac{\mathrm{d}\Delta}{\mathrm{d}\ln\Lambda} \approx \Delta\left(1-\frac{\lambda^2C}{4\pi}\right),\\[10pt]
\displaystyle\frac{\mathrm{d}\lambda}{\mathrm{d}\ln\Lambda} \approx \lambda\left(\frac{1}2-\dfr{\lambda^2 C}{4\pi}\right).
\end{array}
\right.
\end{gather}
The flows of the renormalization group for the ohmic spin-boson model are shown in Fig.\,\ref{S}.
From it, one can see that the theory contains two fixed points: the Gaussian one and the WF point:
\begin{gather*}
\Delta=\Delta^*=0,\quad h=h^*=0,\quad\lambda=\lambda^*=\sqrt{{2\pi/{C}}}.
\end{gather*}

\section*{APPENDIX C}

For deriving the renormalized parameter $\Delta_{min}$ of the delocalized state, one can use the Anderson ``poor man's'' method \cite{RevModPhys.59.1,PhysRevB.9.215,hewson1980local}:
Let us consider the ohmic case. Since the integral in (\ref{R1}) is small, we present the expression for each iteration of the renormalization procedure in exponential form:
\begin{multline}
 \Delta_{n}= \Delta_{n-1}\left[1-\frac{\lambda^2C}{2}\int\limits_{\omega_n}^{\omega_{n-1}}\dfr{\mathrm{d}\omega}{2\pi}\dfr{1}{\omega}\right]\\=
 \Delta_{n-1}\exp\left[-\frac{\lambda^2C}2\int\limits^{\omega_{n-1}}_{\omega_n}\frac{\mathrm{d}\omega}{2\pi}\frac{1}{\omega}\right].
\end{multline}
If $\omega_0=\omega_D$ (Debye frequency) and $\Delta_0=\Delta$ are the initial parameters, then the result of $N$ iterations is
\begin{multline}
\Delta_N=\Delta\exp\left[-\frac{\lambda^2C}2\sum\limits_{n=1}^N\int\limits^{\omega_{n-1}}_{\omega_n}\frac{\mathrm{d}\omega}{2\pi}\frac{1}{\omega}\right]\\=
\Delta\exp\left[-\frac{\lambda^2C}2\int\limits^{\omega_D}_{\omega_N}\frac{\mathrm{d}\omega}{2\pi}\frac{1}{\omega}\right]. 
\end{multline}
The condition that the iteration process stops when $\Delta$ reaches a finite value $\Delta_{min}$ may be expressed as follows:    
\begin{multline}
\Delta_{min}=\Delta\exp\left[-\frac{\lambda^2C}2\int\limits^{\omega_D}_{\Delta_{min}}\frac{\mathrm{d}\omega}{2\pi}\frac{1}{\omega}\right]\\=
\Delta\exp\left[-\frac{\lambda^2C}2\ln\left(\frac{\omega_D}{\Delta_{min}}\right)\right]=
\Delta\left(\frac{\Delta_{min}}{\omega_D}\right)^{\frac{\lambda^2C}{4\pi}} .  \end{multline}
Thus, the renormalized value is
\begin{gather}
 \Delta_{min}= \omega_D\left(\frac{\Delta}{\omega_D}\right)^{\frac{1}{1-\lambda^2C/4\pi}} . 
\end{gather}
This expression is true in the case $\lambda<\lambda_c\equiv\sqrt{2}\lambda^*$ \cite{RevModPhys.59.1,PhysRevB.9.215,hewson1980local}. Thus, the value of the coupling constant $\lambda=\lambda_c$ is the critical one corresponding to the delocalized--localized state transition point.

\section*{APPENDIX D: Bloch--Redfield expression}

Spin commutation relations can (or rather, must) be taken into account when calculating average values. For example, the spin temporal correlation function is represented as follows: \begin{multline}
\langle \sigma_z\sigma_z\rangle_t=-4\langle \sigma_x\sigma_y,\sigma_x\sigma_y\rangle=\\-4\langle \psi_y\psi_z\psi_z\psi_x,\psi_y\psi_z\psi_z\psi_x\rangle_t,
\end{multline}
where the comma in the brackets of the averaging operation divides the operator groups corresponding to different time moments. 
\begin{figure}[h!]
   \centering
\includegraphics[scale=0.4]{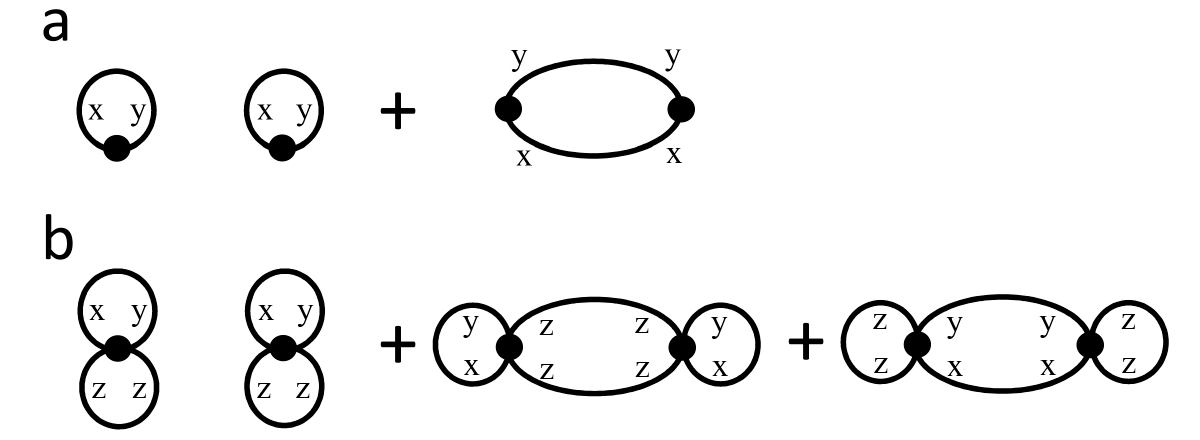}
   \caption{The diagrammatic representation of the correlation function $\langle \sigma_z\sigma_z\rangle_t$: a) which presented in the form $\langle \sigma_z\sigma_z\rangle_t=-\langle {\psi^{cl}_x\psi^{cl}_y},\,\psi^{cl}_x\psi^{cl}_y\rangle_t$; b) which presented in the form $\langle \sigma_z\sigma_z\rangle_t=-4\langle {\psi^{cl}_x\psi^{cl}_y\psi^{cl}_z\psi^{cl}_z},\,{\psi^{cl}_x\psi^{cl}_y\psi^{cl}_z\psi^{cl}_z}\rangle_t$.}
   \label{BRE}
\end{figure}
This expression is presented diagrammatically in Fig.\,\ref{BRE} b.
It is easy to show that in frequency representation, this correlation function can be represented as follows:
\begin{multline*}
\langle \sigma_z\sigma_z\rangle_{\omega}=-4\langle {\psi^{cl}_x\psi^{cl}_y\psi^{cl}_z\psi^{cl}_z},\,{\psi^{cl}_x\psi^{cl}_y\psi^{cl}_z\psi^{cl}_z}\rangle_{\omega}=\\
-8\pi\delta(\omega)\langle \psi^{cl}_x\psi^{cl}_y\rangle\langle\psi^{cl}_z\psi^{cl}_z\rangle,\langle\psi^{cl}_x\psi^{cl}_y\rangle\langle\psi^{cl}_z\psi^{cl}_z\rangle-\\
4\langle\psi^{cl}_z\psi^{cl}_z\rangle\langle\psi^{cl}_z\psi^{cl}_z\rangle\int\frac{\mathrm{d}\omega'}{2\pi}\langle \psi^{cl}_x,\,\psi^{cl}_x\rangle_{\omega'}\langle\psi^{cl}_y,\,\psi^{cl}_y\rangle_{\omega-\omega'}-\\
4\langle\psi^{cl}_x\psi^{cl}_y\rangle\langle\psi^{cl}_x\psi^{cl}_y\rangle\int\frac{\mathrm{d}\omega'}{2\pi}\langle \psi^{cl}_z,\,\psi^{cl}_z\rangle_{\omega'}\langle\psi^{cl}_z,\,\psi^{cl}_z\rangle_{\omega-\omega'}=\\
2\pi\delta(\omega)\langle \sigma_z\rangle^2-\int\frac{\mathrm{d}\omega'}{2\pi}\langle \psi^{cl}_x,\,\psi^{cl}_x\rangle_{\omega'}\langle\psi^{cl}_y,\,\psi^{cl}_y\rangle_{\omega-\omega'}+\\
4\langle \sigma_z\rangle^2\int\frac{\mathrm{d}\omega'}{2\pi}\langle \psi^{cl}_z,\,\psi^{cl}_z\rangle_{\omega'}\langle\psi^{cl}_z,\,\psi^{cl}_z\rangle_{\omega-\omega'}=\\
2\pi\delta(\omega)\langle \sigma_z\rangle^2+\\
\left(1-4\langle \sigma_z\rangle^2\right)\int\frac{\mathrm{d}\omega'}{2\pi}\frac{-\omega'(\omega-\omega')}{((\omega-\omega')^2-\Sigma^2)(\omega'^2-\Sigma^2)}+O=\\
2\pi\delta(\omega)\langle \sigma_z\rangle^2+
\left(1-4\langle \sigma_z\rangle^2\right)\frac{\mathrm{i}\Sigma}{\omega^2-4\Sigma^2}+O,
\end{multline*}
where $O$ is the rapidly decreasing part of the correlation function. 
The obtained representation of the spin correlation function is the well-known Bloch--Redfield expression~\cite{PhysRevB.55.142}, or the so-called ``golden rule'' approach to the spin-boson problem~\cite{RevModPhys.59.1}.

\section*{Acknowledgments}
We are deeply grateful to Y. Makhlin, I. Burmistrov, and M. Gnatich for fruitful discussions regarding this work and for their valuable remarks.

\bibliography{librarySBM}

@book{Kamenev_2011, place={Cambridge}, title={Field Theory of Non-Equilibrium Systems}, publisher={Cambridge University Press}, author={Kamenev, Alex}, year={2011}}

@article{RevModPhys.59.1,
  title = {Dynamics of the dissipative two-state system},
  author = {Leggett, A. J. and Chakravarty, S. and Dorsey, A. T. and Fisher, Matthew P. A. and Garg, Anupam and Zwerger, W.},
  journal = {Rev. Mod. Phys.},
  volume = {59},
  issue = {1},
  pages = {1--85},
  numpages = {0},
  year = {1987},
  month = {Jan},
  publisher = {American Physical Society},
  doi = {10.1103/RevModPhys.59.1},
  url = {https://link.aps.org/doi/10.1103/RevModPhys.59.1}
}

@article{PhysRevB.100.115106,
  title = {Quantum phase transitions in the spin-boson model without the counterrotating terms},
  author = {Wang, Yan-Zhi and He, Shu and Duan, Liwei and Chen, Qing-Hu},
  journal = {Phys. Rev. B},
  volume = {100},
  issue = {11},
  pages = {115106},
  numpages = {9},
  year = {2019},
  month = {Sep},
  publisher = {American Physical Society},
  doi = {10.1103/PhysRevB.100.115106},
  url = {https://link.aps.org/doi/10.1103/PhysRevB.100.115106}
}

@article{PhysRevLett.91.207204,
  title = {Spin-Spin Correlators in the Majorana Representation},
  author = {Shnirman, Alexander and Makhlin, Yuriy},
  journal = {Phys. Rev. Lett.},
  volume = {91},
  issue = {20},
  pages = {207204},
  numpages = {4},
  year = {2003},
  month = {Nov},
  publisher = {American Physical Society},
  doi = {10.1103/PhysRevLett.91.207204},
  url = {https://link.aps.org/doi/10.1103/PhysRevLett.91.207204}
}

@article{PhysRevB.93.174420,
  title = {Using Majorana spin-$\frac{1}{2}$ representation for the spin-boson model},
  author = {Schad, Pablo and Shnirman, Alexander and Makhlin, Yuriy},
  journal = {Phys. Rev. B},
  volume = {93},
  issue = {17},
  pages = {174420},
  numpages = {7},
  year = {2016},
  month = {May},
  publisher = {American Physical Society},
  doi = {10.1103/PhysRevB.93.174420},
  url = {https://link.aps.org/doi/10.1103/PhysRevB.93.174420}
}

@article{PhysRevB.55.142,
  title = {Theory of finite-temperature crossovers near quantum critical points close to,or above, their upper-critical dimension},
  author = {Sachdev, Subir},
  journal = {Phys. Rev. B},
  volume = {55},
  issue = {1},
  pages = {142--163},
  numpages = {0},
  year = {1997},
  month = {Jan},
  publisher = {American Physical Society},
  doi = {10.1103/PhysRevB.55.142},
  url = {https://link.aps.org/doi/10.1103/PhysRevB.55.142}
}

@article{PhysRevLett.91.207203,
  title = {Spin Dynamics from Majorana Fermions},
  author = {Mao, W. and Coleman, P. and Hooley, C. and Langreth, D.},
  journal = {Phys. Rev. Lett.},
  volume = {91},
  issue = {20},
  pages = {207203},
  numpages = {4},
  year = {2003},
  month = {Nov},
  publisher = {American Physical Society},
  doi = {10.1103/PhysRevLett.91.207203},
  url = {https://link.aps.org/doi/10.1103/PhysRevLett.91.207203}
}

@article{PhysRevB.81.121105,
  title = {Quantum phase transition in the sub-Ohmic spin-boson model: An extended coherent-state approach},
  author = {Zhang, Yu-Yu and Chen, Qing-Hu and Wang, Ke-Lin},
  journal = {Phys. Rev. B},
  volume = {81},
  issue = {12},
  pages = {121105},
  numpages = {4},
  year = {2010},
  month = {Mar},
  publisher = {American Physical Society},
  doi = {10.1103/PhysRevB.81.121105},
  url = {https://link.aps.org/doi/10.1103/PhysRevB.81.121105}
}

@article{PhysRevLett.102.030601,
  title = {Quantum Phase Transition in the Sub-Ohmic Spin-Boson Model: Quantum Monte Carlo Study with a Continuous Imaginary Time Cluster Algorithm},
  author = {Winter, Andr\'e and Rieger, Heiko and Vojta, Matthias and Bulla, Ralf},
  journal = {Phys. Rev. Lett.},
  volume = {102},
  issue = {3},
  pages = {030601},
  numpages = {4},
  year = {2009},
  month = {Jan},
  publisher = {American Physical Society},
  doi = {10.1103/PhysRevLett.102.030601},
  url = {https://link.aps.org/doi/10.1103/PhysRevLett.102.030601}
}

@article{PhysRevLett.108.160401,
  title = {Critical and Strong-Coupling Phases in One- and Two-Bath Spin-Boson Models},
  author = {Guo, Cheng and Weichselbaum, Andreas and von Delft, Jan and Vojta, Matthias},
  journal = {Phys. Rev. Lett.},
  volume = {108},
  issue = {16},
  pages = {160401},
  numpages = {5},
  year = {2012},
  month = {Apr},
  publisher = {American Physical Society},
  doi = {10.1103/PhysRevLett.108.160401},
  url = {https://link.aps.org/doi/10.1103/PhysRevLett.108.160401}
}

@article{doi:10.1080/14786430500070396,
author = {Matthias Vojta},
title = {Impurity quantum phase transitions},
journal = {Philosophical Magazine},
volume = {86},
number = {13-14},
pages = {1807--1846},
year = {2006},
publisher = {Taylor \& Francis},
doi = {10.1080/14786430500070396},
URL = {        https://doi.org/10.1080/14786430500070396
},
eprint = {         https://doi.org/10.1080/14786430500070396
}
}

@article{PhysRevLett.69.2142,
  title = {New fermionic description of quantum spin liquid state},
  author = {Tsvelik, A. M.},
  journal = {Phys. Rev. Lett.},
  volume = {69},
  issue = {14},
  pages = {2142--2144},
  numpages = {0},
  year = {1992},
  month = {Oct},
  publisher = {American Physical Society},
  doi = {10.1103/PhysRevLett.69.2142},
  url = {https://link.aps.org/doi/10.1103/PhysRevLett.69.2142}
}

@article{SCHAD2015401,
title = {Majorana representation for dissipative spin systems},
journal = {Annals of Physics},
volume = {361},
pages = {401-422},
year = {2015},
issn = {0003-4916},
doi = {https://doi.org/10.1016/j.aop.2015.07.006},
url = {https://www.sciencedirect.com/science/article/pii/S0003491615002717},
author = {P. Schad and Yu. Makhlin and B.N. Narozhny and G. Schön and A. Shnirman}
}

@article{PhysRevLett.94.070604,
  title = {Quantum Phase Transitions in the Sub-Ohmic Spin-Boson Model: Failure of the Quantum-Classical Mapping},
  author = {Vojta, Matthias and Tong, Ning-Hua and Bulla, Ralf},
  journal = {Phys. Rev. Lett.},
  volume = {94},
  issue = {7},
  pages = {070604},
  numpages = {4},
  year = {2005},
  month = {Feb},
  publisher = {American Physical Society},
  doi = {10.1103/PhysRevLett.94.070604},
  url = {https://link.aps.org/doi/10.1103/PhysRevLett.94.070604}
}

@article{PhysRevB.9.215,
  title = {Low- temperature properties of the Kondo Hamiltonian},
  author = {Emery, V. J. and Luther, A.},
  journal = {Phys. Rev. B},
  volume = {9},
  issue = {1},
  pages = {215--226},
  numpages = {0},
  year = {1974},
  month = {Jan},
  publisher = {American Physical Society},
  doi = {10.1103/PhysRevB.9.215},
  url = {https://link.aps.org/doi/10.1103/PhysRevB.9.215}
}

@book{doi:10.1142/6738,
author = {Weiss, Ulrich},
title = {Quantum Dissipative Systems},
publisher = {WORLD SCIENTIFIC},
year = {2008},
doi = {10.1142/6738},
address = {},
edition   = {3rd},
URL = {https://www.worldscientific.com/doi/abs/10.1142/6738},
eprint = {https://www.worldscientific.com/doi/pdf/10.1142/6738}
}

@article{PhysRevLett.91.170601,
  title = {Numerical Renormalization Group for Bosonic Systems and Application to the Sub-Ohmic Spin-Boson Model},
  author = {Bulla, Ralf and Tong, Ning-Hua and Vojta, Matthias},
  journal = {Phys. Rev. Lett.},
  volume = {91},
  issue = {17},
  pages = {170601},
  numpages = {4},
  year = {2003},
  month = {Oct},
  publisher = {American Physical Society},
  doi = {10.1103/PhysRevLett.91.170601},
  url = {https://link.aps.org/doi/10.1103/PhysRevLett.91.170601}
}

@article{hewson1980local,
  title={On the local polaron model and its applications to intermediate valence systems},
  author={Hewson, AC and Newns, DM},
  journal={Journal of Physics C: Solid State Physics},
  volume={13},
  number={24},
  pages={4477},
  year={1980},
  publisher={IOP Publishing}
}

@article{kosterlitz1976phase,
  title={Phase transitions in long-range ferromagnetic chains},
  author={Kosterlitz, JM},
  journal={Physical Review Letters},
  volume={37},
  number={23},
  pages={1577},
  year={1976},
  publisher={APS}
}

@article{PhysRevLett.18.1049,
  title = {Infrared Catastrophe in Fermi Gases with Local Scattering Potentials},
  author = {Anderson, P. W.},
  journal = {Phys. Rev. Lett.},
  volume = {18},
  issue = {24},
  pages = {1049--1051},
  numpages = {0},
  year = {1967},
  month = {Jun},
  publisher = {American Physical Society},
  doi = {10.1103/PhysRevLett.18.1049},
  url = {https://link.aps.org/doi/10.1103/PhysRevLett.18.1049}
}

@article{PhysRevB.85.144425,
  title = {Scaling analysis in the numerical renormalization group study of the sub-Ohmic spin-boson model},
  author = {Tong, Ning-Hua and Hou, Yan-Hua},
  journal = {Phys. Rev. B},
  volume = {85},
  issue = {14},
  pages = {144425},
  numpages = {14},
  year = {2012},
  month = {Apr},
  publisher = {American Physical Society},
  doi = {10.1103/PhysRevB.85.144425},
  url = {https://link.aps.org/doi/10.1103/PhysRevB.85.144425}
}

@article{PhysRevLett.110.010402,
  title = {Persistence of Coherent Quantum Dynamics at Strong Dissipation},
  author = {Kast, Denis and Ankerhold, Joachim},
  journal = {Phys. Rev. Lett.},
  volume = {110},
  issue = {1},
  pages = {010402},
  numpages = {5},
  year = {2013},
  month = {Jan},
  publisher = {American Physical Society},
  doi = {10.1103/PhysRevLett.110.010402},
  url = {https://link.aps.org/doi/10.1103/PhysRevLett.110.010402}
}

@article{PhysRevB.81.054308,
  title = {Ultraslow quantum dynamics in a sub-Ohmic heat bath},
  author = {Nalbach, P. and Thorwart, M.},
  journal = {Phys. Rev. B},
  volume = {81},
  issue = {5},
  pages = {054308},
  numpages = {7},
  year = {2010},
  month = {Feb},
  publisher = {American Physical Society},
  doi = {10.1103/PhysRevB.81.054308},
  url = {https://link.aps.org/doi/10.1103/PhysRevB.81.054308}
}

@article{PhysRevLett.129.120406,
  title = {Hidden Phase of the Spin-Boson Model},
  author = {Otterpohl, Florian and Nalbach, Peter and Thorwart, Michael},
  journal = {Phys. Rev. Lett.},
  volume = {129},
  issue = {12},
  pages = {120406},
  numpages = {5},
  year = {2022},
  month = {Sep},
  publisher = {American Physical Society},
  doi = {10.1103/PhysRevLett.129.120406},
  url = {https://link.aps.org/doi/10.1103/PhysRevLett.129.120406}
}

@article{shen2023numerical,
  title={Numerical variational studies of quantum phase transitions in the sub-Ohmic spin-boson model with multiple polaron ansatz},
  author={Shen, Yulong and Zhou, Nengji},
  journal={Computer Physics Communications},
  volume={293},
  pages={108895},
  year={2023},
  publisher={Elsevier}
}

@book{Nielsen2000,
  author    = {Nielsen, M. A. and Chuang, I. L.},
  title     = {Quantum Computation and Quantum Information},
  publisher = {Cambridge University Press},
  year      = {2000},
  address   = {Cambridge}
}

@article{RevModPhys.75.715,
  title = {Decoherence, einselection, and the quantum origins of the classical},
  author = {Zurek, Wojciech Hubert},
  journal = {Rev. Mod. Phys.},
  volume = {75},
  issue = {3},
  pages = {715--775},
  numpages = {0},
  year = {2003},
  month = {May},
  publisher = {American Physical Society},
  doi = {10.1103/RevModPhys.75.715},
  url = {https://link.aps.org/doi/10.1103/RevModPhys.75.715}
}

@article{RevModPhys.73.357,
  title = {Quantum-state engineering with Josephson-junction devices},
  author = {Makhlin, Yuriy and Sch\"on, Gerd and Shnirman, Alexander},
  journal = {Rev. Mod. Phys.},
  volume = {73},
  issue = {2},
  pages = {357--400},
  numpages = {0},
  year = {2001},
  month = {May},
  publisher = {American Physical Society},
  doi = {10.1103/RevModPhys.73.357},
  url = {https://link.aps.org/doi/10.1103/RevModPhys.73.357}
}

@article{
doi:10.1126/science.1231930,
author = {M. H. Devoret  and R. J. Schoelkopf },
title = {Superconducting Circuits for Quantum Information: An Outlook},
journal = {Science},
volume = {339},
number = {6124},
pages = {1169-1174},
year = {2013},
doi = {10.1126/science.1231930},
URL = {https://www.science.org/doi/abs/10.1126/science.1231930},
eprint = {https://www.science.org/doi/pdf/10.1126/science.1231930}}

@book{Weiss2012,
author = {Weiss, Ulrich},
year = {2012},
month = {01},
pages = {1-566},
title = {Quantum dissipative systems, fourth edition},
isbn = {978-981-4374-91-0},
doi = {10.1142/8334}
}

@article{Clarke2008,
  author = {Clarke, John and Wilhelm, Frank K.},
  title = {Superconducting quantum bits},
  journal = {Nature},
  year = {2008},
  volume = {453},
  number = {7198},
  pages = {1031--1042},
  month = {June},
  doi = {10.1038/nature07128},
  url = {https://doi.org/10.1038/nature07128},
  issn = {1476-4687}
}

@article{PhysRevB.77.174305,
  title = {Density matrix renormalization group approach to the spin-boson model},
  author = {Wong, Hang and Chen, Zhi-De},
  journal = {Phys. Rev. B},
  volume = {77},
  issue = {17},
  pages = {174305},
  numpages = {6},
  year = {2008},
  month = {May},
  publisher = {American Physical Society},
  doi = {10.1103/PhysRevB.77.174305},
  url = {https://link.aps.org/doi/10.1103/PhysRevB.77.174305}
}

@article{PhysRevLett.102.150601,
  title = {Sparse Polynomial Space Approach to Dissipative Quantum Systems: Application to the Sub-Ohmic Spin-Boson Model},
  author = {Alvermann, A. and Fehske, H.},
  journal = {Phys. Rev. Lett.},
  volume = {102},
  issue = {15},
  pages = {150601},
  numpages = {4},
  year = {2009},
  month = {Apr},
  publisher = {American Physical Society},
  doi = {10.1103/PhysRevLett.102.150601},
  url = {https://link.aps.org/doi/10.1103/PhysRevLett.102.150601}
}

@article{PhysRevB.4.3174,
  title = {Renormalization Group and Critical Phenomena. I. Renormalization Group and the Kadanoff Scaling Picture},
  author = {Wilson, Kenneth G.},
  journal = {Phys. Rev. B},
  volume = {4},
  issue = {9},
  pages = {3174--3183},
  numpages = {0},
  year = {1971},
  month = {Nov},
  publisher = {American Physical Society},
  doi = {10.1103/PhysRevB.4.3174},
  url = {https://link.aps.org/doi/10.1103/PhysRevB.4.3174}
}

@article{PhysRevLett.29.917,
  title = {Critical Exponents for Long-Range Interactions},
  author = {Fisher, Michael E. and Ma, Shang-keng and Nickel, B. G.},
  journal = {Phys. Rev. Lett.},
  volume = {29},
  issue = {14},
  pages = {917--920},
  numpages = {0},
  year = {1972},
  month = {Oct},
  publisher = {American Physical Society},
  doi = {10.1103/PhysRevLett.29.917},
  url = {https://link.aps.org/doi/10.1103/PhysRevLett.29.917}
}

\end{document}